# Snowmass2021 White Paper AF3- CEPC

## CEPC Accelerator Study Group[1]

## 1. Design Overview

### 1.1 Introduction and status

The discovery of the Higgs boson at CERN's Large Hadron Collider (LHC) in July 2012 raised new opportunities for large-scale accelerators. The Higgs boson is the heart of the Standard Model (SM), and is at the center of many biggest mysteries, such as the large hierarchy between the weak scale and the Planck scale, the nature of the electroweak phase transition, the original of mass, the nature of dark matter, the stability of vacuum, etc. and many other related questions. Precise measurements of the properties of the Higgs boson serve as probes of the underlying fundamental physics principles of the SM and beyond. Due to the modest Higgs boson mass of 125 GeV, it is possible to produce it in the relatively clean environment of a circular electron–positron collider with high luminosity, new technologies, low cost, and reduced power consumption. In September 2012, Chinese scientists proposed a 240 GeV *Circular Electron Positron Collider* (CEPC), serving two large detectors for Higgs studies and other topics as shown in Fig. 1. The ~100 km tunnel for such a machine could also host a *Super Proton Proton Collider* (SPPC) to reach energies well beyond the LHC.

The CEPC is a large international scientific project initiated and to be hosted by China. It was presented for the first time to the international community at the ICFA Workshop *"Accelerators for a Higgs Factory: Linear vs. Circular"* (HF2012) in November 2012 at Fermilab. A Preliminary Conceptual Design Report (Pre-CDR, the White Report)[1] was published in March 2015, followed by a Progress Report (the Yellow Report)[2] in April 2017, in which the CEPC accelerator baseline choice was made. The Conceptual Design Report (CEPC Accelerator CDR, *the Blue Report*) [3] has been completed in July 2018 by hundreds of scientists and engineers after an international review from June 28-30, 2018 and was formally released in Nov. 2018. In May 2019, CEPC accelerator document was submitted to European High Energy Physics Strategy workshop for worldwide discussions [4]. After the CEPC CDR, CEPC accelerator entered the phase of Technical Design Report (TDR) endorsed by CEPC International Advisory Committee (IAC). In TDR phase, CEPC optimization design with higher performance compared with CDR and the key technologies such as 650MHz high power and high efficiency klystron, high quality SRF accelerator technology, high precision magnets for booster and collider rings, vacuum system, MDI, etc. have been carried out, and the CEPC accelerator TDR will be completed at the end of 2022. After the TDR, CEPC will enter into Engineering Design Report (EDR) phase, from 2023-2025, with the focusing points on detailed engineering design of accelerator components, industrialization preparation, site selection converging and detailed site geological investigation.

---
[1] Correspondance: J. Gao, Institute of High Energy Physics, CAS, China
Email: gaoj@ihep.ac.cn

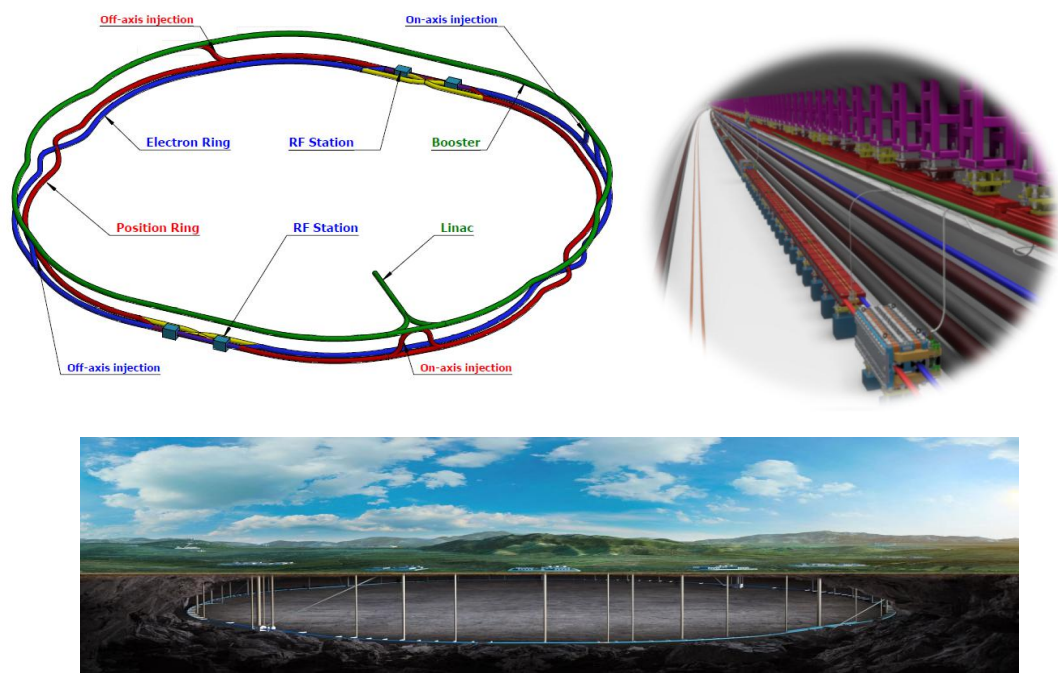

Fig. 1 CEPC layout: Linac injector, booster and collider rings

## 1.2 Performance matrix

- **Energy**

The CEPC is a circular *e+e-* collider Higgs Factory located in a 100 km circumference tunnel beneath the ground. The accelerator complex consists of a linear accelerator (Linac), a damping ring (DR), the full energy injection Booster and the Collider (both booster and collider rings are in the same tunnel) and several connecting transfer lines. The center-of-mass energy of the CEPC is set at 240 GeV, and at that collision energy the CEPC will serve as a Higgs factory, generating millions of Higgs particles. The design also allows for operation at 91 GeV as a Z factory, and at 160 GeV as a W factory, upgradable to ttbar energy of 180GeV. The number of Z particles produced will be more than one trillion, and W⁺W⁻ pairs more than 20 million.

In addition to particle physics, the *e+e-* collider can also operate simultaneously as a powerful synchrotron radiation (SR) light source with photon energy up to 300MeV.

- **Upgrade and staging potential**

CEPC upgrade potentials are multifold. Firstly, the highest energy is at Higgs energy of 240GeV (center of mass), but can be operated at W and Z-pole energies all with 30MW SR power per beam. Secondly, the SR power per beam could be upgraded to 50MW, and finally, CEPC could be upgraded to ttbar energy of 360GeV

(center of mass).

After the CEPC operation period, in the same tunnel, a future *Supper pp Collider*, SPPC, can be accommodated without removing the CEPC collider rings. This opens up the exciting possibilities of *e-p* and *e*-ion physics in addition to *e+e-* physics (CEPC) and pp and ion-ion physics (SPPC).

- **Luminosity**

According to the CEPC TDR design, the circulating CEPC beams radiate 30 MW (upgradable to 50MW) synchrotron radiation power per beam, and the total facility power consumption is kept below 300 MW (upgradable to 500MW). The luminosities (30MW SR power/beam) at the Higgs, W, Z-pole and ttbar energies are $5 \times 10^{34}$ cm$^{-2}$s$^{-1}$, $16 \times 10^{34}$ cm$^{-2}$s$^{-1}$, $115 \times 10^{34}$ cm$^{-2}$s$^{-1}$ and $0.5 \times 10^{34}$ cm$^{-2}$s$^{-1}$ per interaction point, respectively. When beam SR power is upgraded to 50MW, the luminosities (50MW SR power/beam) at the Higgs, W, Z-pole and ttbar energies are $8.3 \times 10^{34}$ cm$^{-2}$s$^{-1}$, $27 \times 10^{34}$ cm$^{-2}$s$^{-1}$, $192 \times 10^{34}$ cm$^{-2}$s$^{-1}$ and $0.8 \times 10^{34}$ cm$^{-2}$s$^{-1}$ per interaction point, respectively.

- **Injector chain**

The heart of the CEPC is a double-ring collider (except at the SCRF region, where electron and positron beams use a common beam pipe). Electron and positron beams circulate in opposite directions in separate beam pipes but with the common SCRF system. They collide at two interaction points (IPs), where two large detectors as described in detail in the CDR (Volume II) are located. The CEPC Booster is located in the same tunnel but above the Colliding rings. It is a synchrotron with a 20 GeV injection energy and extraction energy equal to the beam collision energy. The repetition cycle is 10 seconds. Top-up injection will be used to maintain constant luminosity. The 20 GeV Linac (S-band +C-band), injector to the Booster, built at ground level, accelerates both electrons and positrons. A 1.1 GeV damping ring reduces the positron beam emittance before the positrons are injected to the booster.

- **Facility scale: civil engineering, conventional facilities and power consumption**

Underground structures of CEPC consist of collider ring tunnel (L=100km), experiment halls (2 experiment halls for CEPC, and 2 future experiment halls for SPPC), about 1.4km CEPC linac injector is located on the surface. Surface structures within the collider ring area, such as auxiliary equipment structures, cooling towers, substations and ventilation systems, are located close to the shafts. The total area of surface structures is 140450m$^2$.The total electrical load for physical experiments and general facilities is 270MW. It is proposed to use 220kV for the project, and to have two 220kV central substations (220kV/110kV/10kV) in the project area. The total heat load absorbed by the cooling water system is 213MW during CEPC operation for Higgs physics experiments. There are 16 pump stations at each point of the ring and an additional one for Linac. The air conditioning cold load of the collider ring tunnel is about 6MW. The tunnel is divided into 32 sections. Each section is considered to be independent for the ventilation and primary return air conditioning system. Fire prevention and exhaust systems, hydrant and fire extinguisher systems, and fire detection and fire alarm systems are combined with building fire prevention and evacuation, to minimize fire hazards.

- **CEPC cost, site selection, construction and management**

The cost of CEPC contains mainly three parts, civil engineering, accelerator with two detectors. As cost repartition (without detectors, for example), civil engineering 40%; collider magnets 12%; booster magnets 6%, vacuum system 9%; RF power system 6%; SRF system 4%; mechanical system 5%; instrumentation 4%; cryogenic system 4%; linac injector 2.5%; and others. The CEPC total CDR cost is 5Billion USD including detectors and contingency. TDR cost will be updated.

As for CEPC site selection, the technical criteria are roughly quantified as follows: earthquake intensity less than seven on the Richter scale; earthquake acceleration less than 0.1 g; ground surface-vibration amplitude less than 20 nm at 1–100 Hz; granite bedrock around 50–100 m deep, and others. The site-selection process started in February 2015，preliminary studies of geological conditions for CEPC's potential site locations have been carried out in Qinhuangdao and Xiongan in Hebei province; Huangling county in Shanxi province; Huzhou in Zhejiang province; Changchun in Jilin province and Changsha, in Hunan province, and all these sites satisfies the CEPC construction requirements. An example site location and geological condition are shown in Fig. 2.

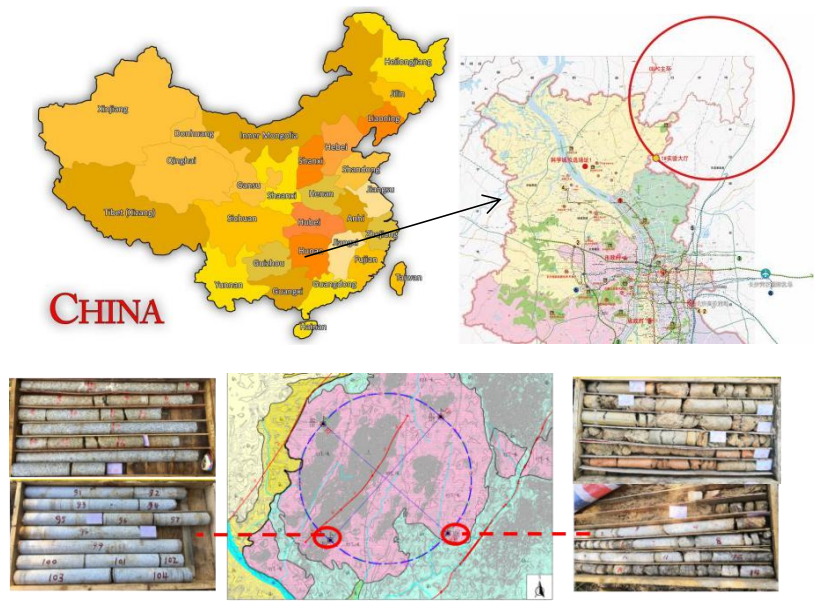

Fig. 2 CEPC Changsha site, Hunan province and geological condition investigation
(one of the site example)

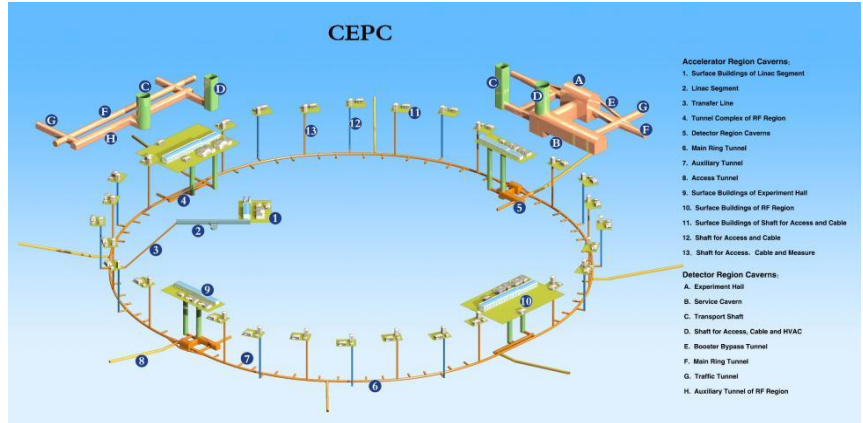

Fig. 3 CEPC tunnel layout

According to Chinese civil construction companies involved in the siting process, a 100 km tunnel will take less than five years to dig using drill-and-blast methods, which is followed by accelerator and detectors' installation. The total CEPC tunnel civil construction period is 54 months, including 8 months for construction preparation, 43 months for construction of main structures and 3 months for completion. The CEPC tunnel layout is shown in Fig. 3.

After the completion of CDR in 2018, prior to the construction, there will be a TDR R&D period to be completed at the end of 2022 and EDR preparation period (2023-2025). During these periods, prototypes of key technical components will be built and infrastructure established for industrialization for manufacturing the large number of required components. Six sites have been considered and they all satisfy the technical requirements. The construction of CEPC is expected to start around 2026 (during the 15th five year plan of China) and will be completed around 2035. After the commissioning, a tentative operation plan will be 7-year running for Higgs physics, followed by a 2-year operation in Z mode and 1-year operation in W mode.

The SppC (Super proton-proton Collider) as an integral part of the CEPC-SppC project, aims at the energy-frontier discoveries, which will be a long-term development or more than twenty years from now after CEPC. As SppC is a long-term project with many technical challenges, it is necessary to pursue studies to work on critical physics and technological key issues. One of the key issues is the iron-based high-field superconducting magnets of at least 20 T to allow proton–proton collisions at a center-of-mass energy of 125 TeV and a luminosity level of $4.3 \times 10^{34}$ cm$^{-2}$s$^{-1}$.

As for the timeline of SppC, from now up to 2035 is CDR and R&D period, and from 2035 to 2045 is the TDR and EDR periods, from 2045-2050 is SppC construction period, and SppC will be put to operation started after 2050.

The CEPC by nature is Chinese initiated international large science project, and the project will be participated, contributed and managed in an international way in all its level and in all of its process from CDR, TDR, construction and operation for physics experiments.

## 1.3 Design summary

### 1.3.1 CEPC TDR parameters

According to the CEPC TDR baseline physics goals at the Higgs and Z-pole energies, the CEPC should provide e+e- collisions at the center-of-mass energy of 240 GeV and deliver a peak luminosity of $5.0 \times 10^{34}$ cm$^{-2}$s$^{-1}$ at each interaction point. The CEPC has two IPs for e+e- collisions and is compatible with four energy modes (tt, Higgs, W and Z-pole). At the Z-pole the luminosity is required to be larger than $1 \times 10^{36}$ cm$^{-2}$s$^{-1}$ per IP. The experiments at ttbar energy is an upgrade at the last stage of CEPC.

The CEPC TDR design is a 100 km double ring scheme based on crab waist collision and 30 MW radiation power per beam at four energy modes, with the shared RF system for Higgs/ttbar energies and independent RF system for W/ Z energies.

The CEPC main parameters for TDR are listed in Table 1. The luminosity at Higgs energy is $5 \times 10^{34}$ cm$^{-2}$s$^{-1}$. At the Z-pole, the luminosity is $1.15 \times 10^{36}$ cm$^{-2}$s$^{-1}$ for 2T detector solenoid. The limit of bunch number at Z-pole comes from the electron cloud instability of the positron beam. A fast transverse feedback system is designed to control the multi-bunch instability induced by impedance at Z-pole.

The crab-waist scheme increases the luminosity by suppressing vertical blow up, which is a must to reach high luminosity. Beamstrahlung is synchrotron radiation excited by the beam-beam force, which is a new phenomenon in a storage ring based collider especially at high energy region. It will increase the energy spread, lengthen the bunch and may reduce the beam lifetime due to the long tail of the photon spectrum. The beam-beam limit at the W/Z is mainly determined by the coherent x-z instability instead of the beamstrahlung lifetime as in the tt/Higgs mode. A smaller phase advance of the FODO cell (60°/60°) for the collider ring optics is chosen at the W/Z mode to supress the beam-beam instability when we consider the beam-beam effect and longitudinal impedance consistently. The CEPC TDR design goals have been evaluated and checked from the point view of beam-beam interaction, which is feasible and achievable.

Table 1: CEPC main parameters in TDR

|  | Higgs | W | Z | ttbar |
|---|---|---|---|---|
| Number of IPs | 2 | | | |
| Circumference [km] | 100 | | | |
| SR power per beam [MW] | 30 | | | |
| Half crossing angle at IP [mrad] | 16.5 | | | |
| Bending radius [km] | 10.7 | | | |
| Energy [GeV] | 120 | 80 | 45.5 | 180 |
| Energy loss per turn [GeV] | 1.8 | 0.357 | 0.037 | 9.1 |
| Piwinski angle | 5.94 | 6.08 | 24.68 | 1.21 |

| Bunch number | 249 | 1297 | 11951 | 35 |
|---|---|---|---|---|
| Bunch spacing [ns] | 636 | 257 | 23 (10% gap) | 4524 |
| Bunch population [$10^{10}$] | 14 | 13.5 | 14 | 20 |
| Beam current [mA] | 16.7 | 84.1 | 803.5 | 3.3 |
| Momentum compaction [$10^{-5}$] | 0.71 | 1.43 | 1.43 | 0.71 |
| Phase advance of arc FODOs [degree] | 90 | 60 | 60 | 90 |
| Beta functions at IP ($\beta$x/$\beta$y) [m/mm] | 0.33/1 | 0.21/1 | 0.13/0.9 | 1.04/2.7 |
| Emittance ($\varepsilon$x/$\varepsilon$y) [nm/pm] | 0.64/1.3 | 0.87/1.7 | 0.27/1.4 | 1.4/4.7 |
| Beam size at IP ($\sigma$x/$\sigma$y) [um/nm] | 15/36 | 13/42 | 6/35 | 39/113 |
| Bunch length (SR/total) [mm] | 2.3/3.9 | 2.5/4.9 | 2.5/8.7 | 2.2/2.9 |
| Energy spread (SR/total) [%] | 0.10/0.17 | 0.07/0.14 | 0.04/0.13 | 0.15/0.20 |
| Energy acceptance (DA/RF) [%] | 1.7/2.2 | 1.2/2.5 | 1.3/1.7 | 2.3/2.6 |
| Beam-beam parameters ($\xi$x/$\xi$y) | 0.015/0.11 | 0.012/0.113 | 0.004/0.127 | 0.071/0.1 |
| RF voltage [GV] | 2.2 (2cell) | 0.7 (2cell) | 0.12 (1cell) | 10 (5cell) |
| RF frequency [MHz] | 650 | | | |
| Beam lifetime [min] | 20 | 55 | 80 | 18 |
| Luminosity per IP[$10^{34}$/cm²/s] | 5.0 | 16.0 | 115.0 | 0.5 |

### 1.3.2 CEPC TDR upgrade parameters

The CEPC TDR upgrade parameters of 50 MW SR power at Higgs, W, Z and ttbar energy operations and the luminosities are shown in Table 2.

Table 2: CEPC main TDR parameters with upgrade

| | Higgs | W | Z | ttbar |
|---|---|---|---|---|
| Number of IPs | 2 | | | |
| Circumference [km] | 100.0 | | | |
| SR power per beam [MW] | 50 | | | |
| Half crossing angle at IP [mrad] | 16.5 | | | |
| Bending radius [km] | 10.7 | | | |
| Energy [GeV] | 120 | 80 | 45.5 | 180 |
| Energy loss per turn [GeV] | 1.8 | 0.357 | 0.037 | 9.1 |
| Piwinski angle | 5.94 | 6.08 | 24.68 | 1.21 |
| Bunch number | 415 | 2162 | 19918 | 58 |

| Bunch spacing [ns] | 385 | 154 | 15 (10% gap) | 2640 |
|---|---|---|---|---|
| Bunch population [$10^{10}$] | 14 | 13.5 | 14 | 20 |
| Beam current [mA] | 27.8 | 140.2 | 1339.2 | 5.5 |
| Momentum compaction [$10^{-5}$] | 0.71 | 1.43 | 1.43 | 0.71 |
| Phase advance of arc FODOs [degree] | 90 | 60 | 60 | 90 |
| Beta functions at IP ($\beta x/\beta y$) [m/mm] | 0.33/1 | 0.21/1 | 0.13/0.9 | 1.04/2.7 |
| Emittance ($\varepsilon x/\varepsilon y$) [nm/pm] | 0.64/1.3 | 0.87/1.7 | 0.27/1.4 | 1.4/4.7 |
| Beam size at IP ($\sigma x/\sigma y$) [um/nm] | 15/36 | 13/42 | 6/35 | 39/113 |
| Bunch length (SR/total) [mm] | 2.3/3.9 | 2.5/4.9 | 2.5/8.7 | 2.2/2.9 |
| Energy spread (SR/total) [%] | 0.10/0.17 | 0.07/0.14 | 0.04/0.13 | 0.15/0.20 |
| Energy acceptance (DA/RF) [%] | 1.7/2.2 | 1.2/2.5 | 1.3/1.7 | 2.3/2.6 |
| Beam-beam parameters ($\xi x/\xi y$) | 0.015/0.11 | 0.012/0.113 | 0.004/0.127 | 0.071/0.1 |
| RF voltage [GV] | 2.2 (2cell) | 0.7 (2cell) | 0.12 (1cell) | 10 (5cell) |
| RF frequency [MHz] | 650 | | | |
| Beam lifetime [min] | 20 | 55 | 80 | 18 |
| Luminosity per IP[$10^{34}/cm^2/s$] | 8.3 | 26.6 | 192 | 0.8 |

## 1.4 Design challenges

### 1.4.1 CEPC collider lattice optics

The CEPC lattice optics is designed with requirements and constraints mainly from top-level parameters, geometry, minimizing cost, compatibility of Higgs, W, Z and ttbar modes, and compatibility with SppC [3,5,6].

The interaction region is designed to provide strong focusing and crab-waist collision[7]. To get a robust lattice, the length from IP to the final strong focusing quadrupole is chosen to be 1.9 m. A large full crossing angle is 33 mrad is chosen to provide large Piwinski angle with constraints from the machine and detector interface. A local chromaticity correction scheme is adopted to get a large momentum acceptance. An asymmetric lattice is adopted to allow softer synchrotron-radiation photons from the upstream part of the IP[8].

For the arc region，twin-aperture dipoles and quadrupoles [3, 9] are used in the arc region to reduce power. The two beams are separated by 35 cm. The FODO cell structure is chosen to provide a large filling factor of dipoles. The non-interleaved-sextupole scheme is selected due to aberration cancellation. For the Higgs and ttbar energy running, the 90/90-degrees phase advances are chosen to balance the aberration cancellation and magnets number. For the W and Z energy running, 60/60-degrees phase advances are chosen to suppress the microwave and transverse mode coupling instability and to increase stable tune area considering the beam-beam effect and longitudinal impedance [10].

In the RF region, the RF cavities are shared by the two rings. Each RF station is divided into two sections for bypassing half of the cavities when running in W or Z modes [5,11]. An electrostatic separator combined with a dipole magnet avoids bending the incoming beam [8]. The sawtooth effect is expected to be curable by tapering the magnet strength to take into account the beam energy at each magnet. The vertical emittance due to the solenoid field coupling is limited and acceptable. The beam optics of the four energy modes are shown in the Fig. 4(a).

The requirements of dynamic aperture (DA) are got from injection and beam-beam effects to get efficient injection and adequate beam life time. A differential evolution algorithm based optimization code has been developed for CEPC, which is a multi-objective code called MODE [12]. The SAD code is used to do the optics calculation and dynamic aperture tracking. Strong synchrotron radiation causes strong radiation damping which helps enlarge the dynamic aperture to some extent. Quantum fluctuations in the synchrotron radiation are considered in SAD, where the random diffusion due to synchrotron radiation in the particle tracking is implemented in each magnet. Totally, 256 arc sextupole families, 8 IR sextupole families, 4 IR multipoles and 8 phase advance tuning knobs between different sections can be used to optimize the DA. The error effects were studied with misalignment and main field error for the magnets. 100 μm are used for the transverse misalignment. The closed orbit correction, dispersion free steering, beta-beating correction are made to cure the error effects [13]. After the corrections, the vertical emittance growth and dynamic aperture are promising.The transverse DA with erros at Higgs energy is shown in Fig. 4(b), which satisfy the design requirement.

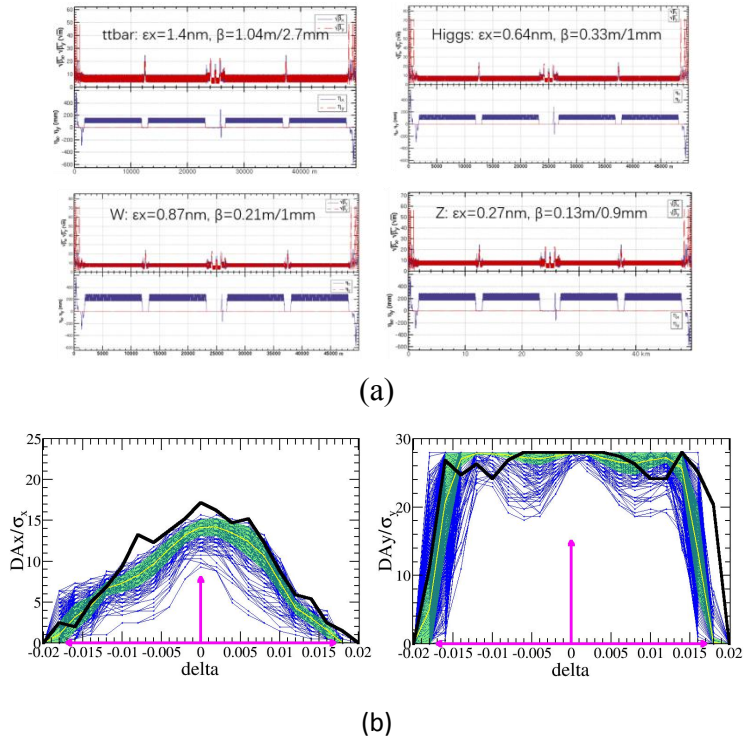

(a)

(b)

Fig. 4 Beam optics for the four energies of CEPC collider ring(a)
and the DA satisfiying the design requirment at Higgs energy with erros (b)

## 1.4.2 Machine-detector interface (MDI)

The machine-detector interface (MDI) issues are one of the most complicate and challenging topics at the Circular Electron Positron Collider (CEPC) and other future high energy colliders. The MDI region is about 14m(±7m from the IP) in length in the Interaction Region (IR), where many elements from both detector system and accelerator components need to be installed including the detector solenoid, anti-solenoid, luminosity calorimeter (LumiCal), interaction region beam pipe, cryostat, beam position monitors (BPMs) and bellows. The cryostat includes the final doublet superconducting magnets and anti-solenoid. The CEPC detector consists of a cylindrical drift chamber surrounded by an electromagnetic calorimeter, which is immersed in a 2~3T superconducting solenoid of 7.6 m in length. The accelerator components inside the detector should not interfere with the devices of the detector. The smaller the conical space occupied by accelerator components, the better will be the geometrical acceptance of the detector. From the requirement of detector, the conical space with an opening angle should not larger than 8.11 degrees. After optimization, the accelerator components inside the detector without shielding are within a conical space with an opening angle of 6.78 degrees. The crossing angle between electron and positron beams is 33 mrad in horizontal plane. The final focusing quadrupole is 1.9 m (L$^*$) from the IP. Primary results are got from the assembly, interfaces with the detector hardware, cooling channels, vibration control of the cryostats, supports and so on.

A water cooling structure is required to control the heating problem of HOM in IR vacuum chamber. The diameters of beryllium pipe and the SC quadrupoles are 20mm.

SR photons in the IR are mainly generated from the final upstream bending magnet and the IR quadrupole magnets due to eccentric particles. With 2 mask tips along the inside of the beam pipe to shadow the inner surface of the pipe the number of scattered photons that can hit the central beam pipe is greatly reduced to only those photons which forward scatter through the mask tips. With collimators in the ARC far from IP the SR photons from IR quadrupoles will not damage the detector components and cause background to experiments.

Beam loss background are mainly from Bhabha scattering, beamstrahlung, beam-thermal photon scattering and beam-gas inelastic scattering. With collimators in the ARC far from IP, beam loss background can be reduced significantly and can be accepted by the detector. The layout of CEPC MDI is shown in Fig. 5.

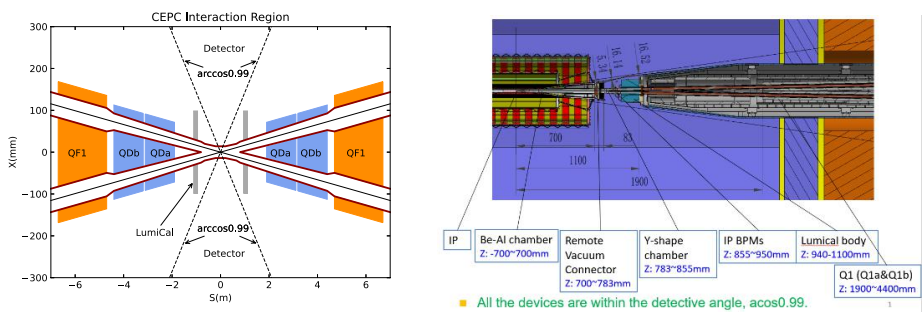

Fig. 5 CEPC MDI layout

### 1.4.3 CEPC booster

The booster provides electron and positron beams to the collider at different energies. The newest booster design is consistent with the TDR higher luminosity goals for four energy modes. The booster is in the same tunnel as the collider, placed above the collider ring except in the interaction region where there are bypasses to avoid the detectors.

The injection system consists of a 20 GeV Linac, followed by a full-energy booster ring. Electron and positron beams are generated and accelerated to 20 GeV in the Linac. The beams are then accelerated to full-energy in the booster, and injected into the collider. For different beam energies of Higgs, W, Z and ttbar, experiments, there will be different particle bunch structures in the collider. To maximize the integrated luminosity, the injection system will operate mostly in top-up mode, and also has the ability to fill the collider from empty to full charge in a reasonable length of time. A traditional off-axis injection scheme is chosen as a baseline design of the beam injection to the collider, and a swap-out injection is given as another choice for Higgs/ttbar injection.

The optics of booster is changed to TME structure and the emittance of booster is reduced significantly in order to match the lower emittance of collider and hence to reach higher luminosity goal in TDR. The length of the TME cell is 80 m and the combined dipole magnets (B+S) are proposed to minimize the cost for magnets and power supplies. The horizontal position of the booster has been designed between the collider two beams and the height difference between booster and collider is 2.4m. CEPC booster has exactly same circumference as the collider which is the requirement of injection scheme and timing system.

During ramping, parasitic sextupole field is induced on beam pipe inside dipoles due to eddy current and the ramping rate is limited by eddy current effect. The ramping time is 4.5 s for Higgs, 2.7 s for W, 1.6 s for Z and 7.2 s for ttbar. An aluminum beam pipe with 55 inner diameter is chosen to make a balance between impedance and eddy effect. The most serious case for eddy current effect is at 30 GeV and this effect has been checked, which is acceptable. The lowest field of dipole magnets is 62 Gauss and the requirement of field error at 20 GeV can be realized by the iron based magnets. The damping time is too long to damp the multi-bunch instability at 20 GeV and the fast feedback systems at low energy region are essential to suppress the instability. The booster parameters at injection and extraction energies are shown in Tables 3 and 4.

Table 3: Main parameters for the booster at injection energy

|  |  | tt | H | W | Z | |
| --- | --- | --- | --- | --- | --- | --- |
| Beam energy | GeV | 20 | | | | |
| Bunch number |  | 37 | 240 | 1230 | 3840 | 5760 |
| Threshold of single bunch current | μA | 5.79 | 4.20 | 3.92 | | |

| Threshold of beam current (limited by coupled bunch instability) | mA | 27 | | | | |
|---|---|---|---|---|---|---|
| Bunch charge | nC | 1.07 | 0.78 | 0.81 | 0.89 | 0.92 |
| Single bunch current | μA | 3.2 | 2.3 | 2.4 | 2.7 | 2.78 |
| Beam current | mA | 0.12 | 0.56 | 2.99 | 10.3 | 16.0 |
| Growth time (coupled bunch instability) | ms | 1690 | 358 | 67 | 19.4 | 12.5 |
| Energy spread | % | 0.016 | | | | |
| Synchrotron radiation loss/turn | MeV | 1.3 | | | | |
| Momentum compaction factor | $10^{-5}$ | 1.12 | | | | |
| Emittance | nm | 0.035 | | | | |
| Natural chromaticity | H/V | -372/-269 | | | | |
| RF voltage | MV | 438.0 | 230.2 | 200.0 | | |
| Betatron tune $\nu_x/\nu_y$ | | 321.23/117.18 | | | | |
| Longitudinal tune | | 0.13 | 0.0943 | 0.0879 | | |
| RF energy acceptance | % | 5.4 | 3.7 | 3.6 | | |
| Damping time | s | 10.4 | | | | |
| Bunch length of linac beam | mm | 0.5 | | | | |
| Energy spread of linac beam | % | 0.16 | | | | |
| Emittance of linac beam | nm | 10 | | | | |

Table 4: Main parameters for the booster at extraction energy

| | | tt | H | | W | Z | |
|---|---|---|---|---|---|---|---|
| | | Off axis injection | Off axis injection | On axis injection | Off axis injection | Off axis injection | |
| Beam energy | GeV | 180 | 120 | | 80 | 45.5 | |
| Bunch number | | 37 | 240 | 233+7 | 1230 | 3840 | 5760 |
| Maximum bunch charge | nC | 0.96 | 0.7 | 23.2 | 0.73 | 0.8 | 0.83 |

| Parameter | Unit | | | | | | |
|---|---|---|---|---|---|---|---|
| Maximum single bunch current | μA | 2.9 | 2.1 | 69.7 | 2.2 | 2.4 | 2.5 |
| Threshold of single bunch current | μA | 91.5 | 70 | | 22.16 | 9.57 | |
| Threshold of beam current (limited by RF system) | mA | 0.3 | 1 | | 4 | 10 | 16 |
| Beam current | mA | 0.11 | 0.51 | 0.99 | 2.69 | 9.2 | 14.4 |
| Growth time (coupled bunch instability) | ms | 16611 | 2359 | 1215 | 297.8 | 49.5 | 31.6 |
| Bunches per pulse of Linac | | 1 | 1 | | 1 | 2 | |
| Time for ramping up | s | 7.3 | 4.5 | | 2.7 | 1.6 | |
| Injection duration for top-up (Both beams) | s | 30.0 | 23.3 | 32.8 | 39.3 | 134.7 | 128.2 |
| Injection interval for top-up | s | 65 | 38 | | 155 | 153.5 | |
| Current decay during injection interval | | 3% | | | | | |
| Energy spread | % | 0.15 | 0.099 | | 0.066 | 0.037 | |
| Synchrotron radiation loss/turn | GeV | 8.45 | 1.69 | | 0.33 | 0.034 | |
| Momentum compaction factor | $10^{-5}$ | 1.12 | | | | | |
| Emittance | nm | 2.83 | 1.26 | | 0.56 | 0.19 | |
| Natural chromaticity | H/V | -372/-269 | | | | | |
| Betatron tune $\nu_x/\nu_y$ | | 321.27/117.19 | | | | | |
| RF voltage | GV | 9.3 | 2.17 | | 0.87 | 0.46 | |
| Longitudinal tune | | 0.13 | 0.0943 | | 0.0879 | 0.0879 | |
| RF energy acceptance | % | 1.34 | 1.59 | | 2.6 | 3.4 | |
| Damping time | ms | 14.2 | 47.6 | | 160.8 | 879 | |
| Natural bunch length | mm | 2.0 | 1.85 | | 1.3 | 0.75 | |
| Full injection from empty ring | h | 0.1 | 0.14 | 0.16 | 0.27 | 1.8 | 0.8 |

### 1.4.4 CEPC linac Injector

The CEPC linac injector is a normal conducting S-band and C-band linac with frequency of 2860 MHz and 5720MHz, providing electron and positron beams at an energy of up to 20 GeV at a repetition rate of 100 Hz, as shown in Fig. 6. One-bunch-per-pulse is adopted and bunch charge is 1.5nC. To keep the potential to meet higher requirements and possibility of updates in the future, the linac can provide electron beam and posirton beam with bunch charge larger than 3.0nC as shown in Table 5. The parameters of S-band and C-band accelerating structures are shown in Table 6. A thermionic electron gun is adopted and can provide 11 nC electron beam for positron production. There is an RF gun scheme as an alternative.The positron source is a conventional design with a tungsten target of 15 mm in length and adiabatic matching device of 6 T in peak magnetic field. The energy of electron beam for positron production is 4 GeV and rms beam size is 0.5 mm.

In the design of CEPC linac, the reliability and availability of the linac injector was emphasized because it is one of the indispensable facilities. The linac has a robust design based on well proven technologies, and about 15% backup of accelerating structure and klystron is foreseen to reach high availability. The linac injector design reached the design requirement as shown in Table 7. The linear type layout of CEPC linac is adopted and one electron transport line in vertical plane at an energy of 1.1 GeV is desiged to bypass the positron source and part accelerating section. A 1.1 GeV damping ring with 147m circumference is adopted to reduce the transverse emittance of positron beam to suitably small value with the parameters shown in Table 8. S-band accelerating structures are used before the damping ring, including a 4GeV electron beam accelerating section and a 1.1GeV positron beam accelerating section. After the damping ring the C-band accelerating structures are used to accelerate electron and positron beam from 1.1GeV to 20GeV. In order to match the beam bunch length to the C-band RF structure, a bunch compressor system is designed. All the lattice design and multi-particle simulations are conducted and the linac with errors can meet all the requirements.

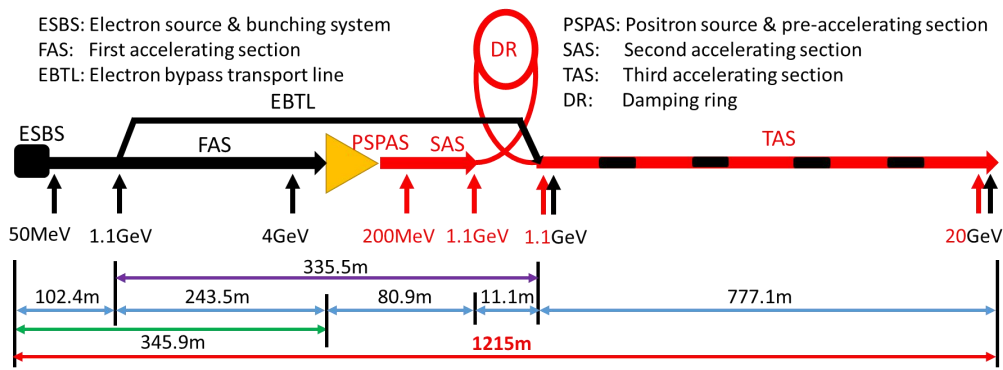

Fig. 6 CEPC linac injector layout

Table 5: CEPC linac injector parameter

| Parameter | Symbol | Unit | Baseline |
|---|---|---|---|
| e- /e+ beam energy | Ee-/Ee+ | GeV | 20 |
| Repetition rate | $f_{rep}$ | Hz | 100 |
| e- /e+ bunch population | Ne-/Ne+ | $\times 10^{10}$ | 0.94(1.88) |
| | | nC | 1.5 (3) |
| Energy spread (e- /e+ ) | $\sigma_E$ | | $1.5 \times 10^{-3}$ |
| Emittance (e- /e+ ) | $\varepsilon_{x,y}$ | nm | 10 |

Table 6: CEPC S-band and C-band linac parameters

| Parameter | Unit | S-band | C-band |
|---|---|---|---|
| Frequency | MHz | 2860 | 5720 |
| Length | m | 3.1 | 1.8 |
| Cavity mode | | $2\pi/3$ | $3\pi/4$ |
| Aperture diameter | mm | 20~24 | 11.8~16 |
| Gradient | MV/m | 21 | 45 |

Table 7: CEPC linac injector achieved parameters

| Parameter | Unit | Baseline | Electron | Positron |
|---|---|---|---|---|
| e- /e+ beam energy | GeV | 20 | 20.38 | 20.37 |
| Repetition rate | Hz | 100 | 100 | 100 |
| e- /e+ bunch population | $\times 10^{10}$ | 0.94(1.88) | 1.88 | 1.88 |
| | nC | 1.5 (3) | 3 | 3 |
| Energy spread (e- /e+ ) | | $1.5 \times 10^{-3}$ | $1.3 \times 10^{-3}$ | $1.5 \times 10^{-3}$ |
| Emittance (e- /e+ ) | nm | 10 | 3.1 | 6.1 |

Table 8: CEPC damping ring parameters

| | DR V3.0 |
|---|---|
| Energy (Gev) | 1.1 |
| Circumference (m) | 147 |
| Number of trains | 2 (4) |
| Number of bunches/trian | 2 |
| Total current (mA) | 12.4 (24.8) |
| Bending radius (m) | 2.87 |
| Dipole strength $B_0$ (T) | 1.28 |
| $U_0$ (kev/turn) | 94.6 |
| Damping time x/y/z (ms) | 11.4/11.4/5.7 |

| Phase/cell (degree) | 60/60 | 75/75 |
|---|---|---|
| Momentum compaction | 0.013 | 0.0075 |
| Storage time (ms) | 20 (40) | |
| $\delta_0$ (%) | 0.056 | |
| $\varepsilon_0$ (mm.mrad) | 94.4 | 56.7 |
| injection $\sigma_z$ (mm) | 5 | 5 |
| Extract $\sigma_z$ (mm) | 4.9 | 4.2 |
| $\varepsilon_{inj}$ (mm.mrad) | 2500 | 2500 |
| $\varepsilon_{ext\ x/y}$ (mm.mrad) | 166(97)/75(3) | 123(59)/68 (2) |
| $\delta_{inj} / \delta_{ext}$ (%) | 0.22 /0.056 | 0.22 /0.056 |
| Energy acceptance by RF(%) | 1.6 | 1.8 |
| $f_{RF}$ (MHz) | 650 | |
| $V_{RF}$ (MV) | 2.0 | 1.5 |
| Longitudinal tune | 0.0346 | 0.0224 |

### 1.4.5 CEPC SRF system

CEPC will use 650 MHz SCRF system for the Collider and 1.3 GHz for the Booster, with the SRF parameters for collider and booster in Tables 9 and 10, respectively. The SRF layout is shown in Fig. 7. RF staging and bypass scheme is proposed to unleash full potential of CEPC to reach highest luminosity at each energy and keep operational flexibility in the same time. RF staging of both the Collider and Booster is required for the CEPC power and energy upgrade. Cavity by-pass is needed to enable seamless operation mode switching, which is different from FCC-ee design. Besides these special requirements, the large scale and wide parameters range of CEPC, i.e. beam energy from 45.5 GeV to 180 GeV, RF voltage from 120 MV to 10 GV, beam current from 3 mA to 1.4 A, SR power per beam up to 50 MW, make its SRF system one of the most challenging accelerator RF systems in history in terms of high Q and high gradient cavities, high power variable input coupler, high power HOM damping, and fundamental mode instability suppression etc. HOM power limit per cavity and the fast-growing longitudinal coupled-bunch instabilities (CBI) driven by both the fundamental and higher order modes impedance of the RF cavities determine to a large extent the highest beam current and luminosity obtainable in the Z mode. Transient beam loading is also a concern. These challenges require intensive R&D and verification as well as new ideas and technologies to push the frontier of superconducting RF.

For the first phase, CEPC will use 240 650 MHz 2-cell superconducting cavities for the Collider and 96 1.3 GHz 9-cell superconducting cavities for the Booster. The Collider is a fully partial double-ring with common cavities for electron and positron beams in Higgs operation mode and a double ring for separate cavities for electron and positron beams in W and Z operation mode. The Collider SRF system is optimized for the Higgs mode of 30 MW SR power per beam as the first priority, with enough tunnel space and operating margin to allow higher RF voltage (ttbar) and SR power (50 MW SR power per beam) by adding cavities. For the high luminosity Z-pole upgrade, dedicated high current Z-pole cavities will be added to by-pass low current Higgs cavities in both the Collider and Booster. Because of the high HOM power and the need to have the smallest number of cavities, KEKB / BEPCII type single cavity cryomodules with very high input coupler power will be needed.

Table 9: CEPC TDR SRF parameters of collider ring

| | ttbar | | Higgs | W | Z |
|---|---|---|---|---|---|
| | Additional 5-cell cavities | Existing 2-cell cavities | | | |
| Luminosity / IP [$10^{34}$ cm$^{-2}$s$^{-1}$] | 0.5 | | 5 | 16 | 115 |
| RF voltage [GV] | 10 (7.8 + 2.2) | | 2.2 | 0.7 | 0.12 |
| Beam current / beam [mA] | 3.4 | | 16.4 | 84 | 803 |
| Bunch charge [nC] | 32 | | 22 | 21.6 | 22.4 |
| Bunch length [mm] | 2.9 | | 3.9 | 4.9 | 8.7 |
| 650 MHz cavity number | 240 | 240 | 240 | 120/ring | 30/ring |
| Cell number / cavity | 5 | 2 | 2 | 2 | 1 |
| Gradient [MV/m] | 28.5 | 20 | 20 | 12.7 | 8.7 |
| $Q_0$ @ 2 K at operating gradient | 5E10 | 2E10 | | | |
| HOM power / cavity [kW] | 0.4 | 0.16 | 0.45 | 0.93 | 2.9 |
| Input power / cavity [kW] | 194 | 56 | 250 | 250 | 1000 |
| Optimal $Q_L$ | 1E7 | 7E6 | 1.6E6 | 6.4E5 | 7.5E4 |
| Optimal detuning [kHz] | 0.01 | 0.02 | 0.1 | 0.9 | 13.3 |
| Cavity number / klystron | 4 | 12 | 2 | 2 | 1 |
| Klystron power [kW] | 1400 | 1400 | 800 | 800 | 1400 |
| Klystron number | 60 | 20 | 120 | 60 | 60 |
| Cavity number / cryomodule | 4 | 6 | | | 1 |
| Cryomodule number | 60 | 40 | | | 30 |
| Total cavity wall loss @ 2 K [kW] | 9.5 | 4.7 | | 1.9 | 0.45 |

Table 10: CEPC TDR SRF parameters of booster ring

| | ttbar | Higgs off/on-axis | W | Z high current |
|---|---|---|---|---|
| Extraction beam energy [GeV] | 180 | 120 | 80 | 45.5 |
| Extraction average SR power [MW] | 0.087 | 0.09 | 0.01 | 0.004 |
| Bunch charge [nC] | 0.96 | 0.7 | 0.73 | 0.83 |
| Beam current [mA] | 0.11 | 0.5/1 | 2.7 | 14.3 |
| Injection RF voltage [GV] | 0.438 | 0.197 | 0.122 | 0.122 |
| Extraction RF voltage [GV] | 9.3 | 2.05 | 0.59 | 0.28 |
| Extraction bunch length [mm] | 1.9 | 1.9 | 1.6 | 0.9 |
| Cavity number (1.3 GHz 9-cell) | 336 | 96 | 64 | 16 |
| Extraction gradient [MV/m] | 26.7 | 20.6 | 8.9 | 17.1 |
| $Q_0$ @ 2 K at operating gradient | 1E10 | | | |
| $Q_L$ | 4E7 | 1E7 | | |
| Cavity bandwidth [Hz] | 33 | 130 | | |
| Peak HOM power per cavity [W] | 0.4 | 1.2/2.3 | 7.8 | 105 |
| Input peak power per cavity [kW] | 7.5 | 16/21.4 | 15 | 31 |
| SSA peak power [kW] (one cavity per SSA) | 10 | 25 | 25 | 40 |
| Cryomodule number (8 cavities per module) | 42 | 12 | 8 | 2 |

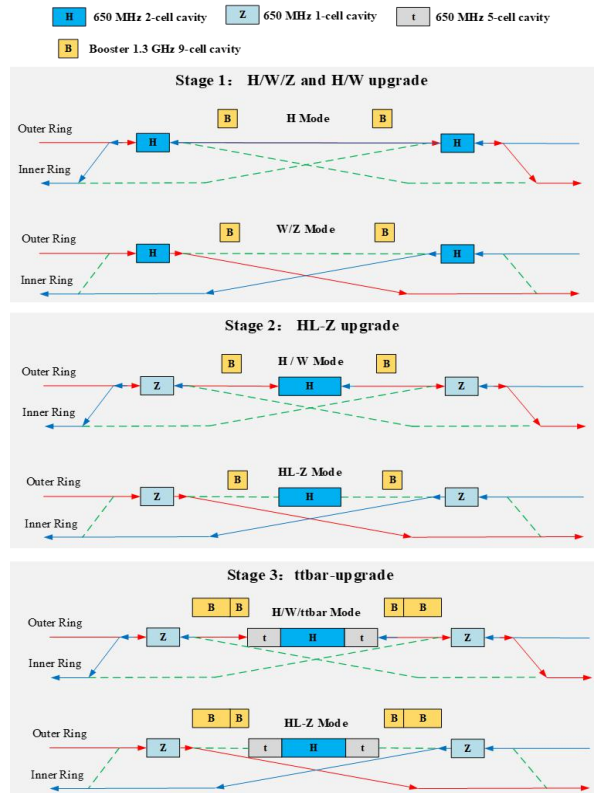

Fig.7 CEPC SRF layout

### 1.4.6 Collective effects

The collective instabilities induced by the interaction of the beam with the vacuum chamber surroundings, or with the electron cloud and residual gas in the vacuum, can induce beam quality degradation or beam loss, and subsequently restrict the machine performance.

A detailed impedance model based on the round beam pipe (ϕ56mm), including resistive wall, RF cavities, flanges, bellows, gate valves, vacuum pumps, BPMs, IR collimators, electro separators, IP chambers, and vacuum transitions, has been developed. The impedance model will be continuously updated to include more contributors. The broadband impedance budget is given and shows that both longitudinal and transverse broadband impedances are dominated by the resistive wall and the elements with large quantities, such as the flanges and bellows. On the other hand, the loss factor is mainly contributed by the resistive wall, RF cavities and bellows.

Based on the impedance model, systematic studies on the collective effects have been carried out. The main constraints from the collective effects are focused on the Z-pole operation mode, due to its high beam current and bunch intensity, as well as slow radiation damping. For the high energy cases, the only concern is that the longitudinal broadband impedance is above the Boussard-Keil-Schnell criterion, which may induce degradation of the beam-beam interaction. More consistent simulation studies with both beam-beam and longitudinal impedance will be performed. For the Z-pole mode, since the beam will be injected with collision, the bunch will be apparently lengthened by both impedance and beamstruhlung. The single bunch intensity will not be constrained by the impedances. However, the total beam current to reach high luminosity is restricted by both the electron cloud effect and the transverse resistive wall instability. In order to reach the luminosity of $\sim 10^{36}/\text{cm}^2/\text{s}$ with SR power per beam of 30 MW, a robust bunch-by-bunch feedback system is required to damp the resistive wall instability with growth time of approximately 2 ms. To reach approximately a factor of 2 higher of the luminosity with SR power per beam of 50 MW, additional mode feedback dampers are required to mitigate the resistive wall instability, meanwhile, antechambers maybe necessary to damp the electron cloud effects.

### 1.4.7 Injection & extraction transfer lines

The CEPC is a circular e+e- collider with a 100-km circumference. It consists of a double-ring collider, a booster, and a linac, and a damping ring to reduce the emittance of the positron beam. The electron and positron beams eventually enter the main collider through acceleration and damping, where they are stored and collided. Dumps are needed in the collider for machine safety. Transport lines are needed to connect these different parts of CEPC.

The transport line from the Linac to the damping ring and from the damping ring back to the Linac requires special design to meet the injection requirements of the damping ring. Before damping ring, energy spread of the beam should be reduced in order to match the RF acceptance of the damping ring. After damping ring, bunch length of the beam should be reduced to control the energy spread after acceleration.

The booster is located between the Linac and the main collision ring. The beam at the exit of the Linac needs to be injected into the booster for acceleration, and the accelerated beam in the booster needs to be extracted and then injected into the main collision ring. The injection and extraction of booster adopts the design of a conventional on-axis injection and extraction. The transport line from the Linac to the booster needs to separate the positron and electron beams and inject them into the booster separately. In order to reduce the cost, the Linac is placed on the ground and the booster is placed in the tunnel 100 meters underground. The transport line also needs to deflect the beam vertically to transport the beam on the ground to the underground tunnel. Therefore, the transport line from the Linac to the booster consists of two parts: vertical deflection section and horizontal deflection section.

The booster-to-collider transport line is used to transport the beam to the main collider. The booster and the main collider are placed in the same tunnel, at a vertical distance of 2.4 meters from the plane of the main collider. The transport line is divided into three sections: the first and third sections are used for horizontal deflection, and the second section is used for vertical deflection.

There are two dumps for each of the electron and positron ring. In the dump transfert lines, one kicker and one septum are used to get the beams into the dump line, so all bunches can be dumped in one turn. Horizontal and vertical dilution kickers are used to change the position of different bunches at the dump, in order to reduce the beam damage to the dump.

## 1.4.8 CEPC timing and bunch patterns

CEPC is a 100 km double ring collider with two interaction points proposed to be working at four energy schemes of Z pole, W, Higgs and ttbar. To achieve the designed luminosity in TDR, the philosophy of both top up injection and full injection from empty for the collider ring is studied and attainable. In addition, the hardwares of injection and extraction for each subsystem are designed and are compatible with different energy modes. According to the requirements of the collider, the injection chain, the injection scheme and the timing structure of CEPC for four energy modes has been designed.

With different beam energies of ttbar, Higgs, W, and Z experiments, there will be different bunch structures in the subsystems. At ttbar and Higgs mode, all the bunches are distributed in half ring for the booster and the collider due to the shared RF scheme, with 4.5us bunch seperation for ttbar and 636ns bunch seperation for Higgs. While the bunches are distributed in the whole ring with 257ns bunch seperation for W and 23ns bunch seperation for Z. The injection scheme and bunch structure at Z pole is designed to be train by train while it is bunch by bunch for other three modes.

The frequency choice was done carefully since CDR with the consideration of physics requirements and timing system. The RF frequency of Linac, damping ring, booster and collider ring is 2860MHz, 650MHz, 1300MHz and 650MHz. The biggest challenge for timing system is the Z-ploe mode, which has about 10,000 to 20,000 bunches and the bunch spacing is about 15ns to 30ns. The common frequency is choosen as 130MHz.

### 1.4.9 Polarization options

Operation of polarized beams at Z-pole and W threshold are under study. Firstly, resonant depolarization technique (RD) using transversely polarized e+ and e- beams are essential for precision measurements of mass & widths of Z and W bosons. To this end, we plan to inject about 150 non-colliding e+/e- bunches, and conduct RD on one bunch every 12 min, to continuously monitor the evolution of center of mass energies. Since the polarization build-up time is 250 hours in the collider ring at 45 GeV, asymmetric wigglers are added to the lattice to boost the initial polarization build-up, about 10% beam polarization can be achieved in 2.6 hours with these wigglers, then these wigglers are turned off to avoid influence on colliding beam experiments. Conceptual designs of transverse Compton polarimeter and depolarizer are also under way. At W threshold, the polarization build-up time is about 15 hours, asymmetric wigglers are not needed. We plan to inject 12 non-colliding e+/e- bunches, and conduct RD on one bunch every 10 min, to continuously monitor of evolution of center of mass energies. The RD technique is itself nontrivial at W energy, with the increasing influence of synchrotron sideband spin resonances, how to properly conduct the RD measurement is also to be studied.

Secondly, longitudinally polarized colliding beam experiments are beneficial for precision electroweak measurements and probe for new physics. A conceptual design of longitudinal beam polarization at Z-pole energy is under way, based on the lattices and parameters of CEPC Conceptual Design Report. The concept involves electron bunches with beam polarization > 80% generated from the source, transported in the injector and injected into the collider ring in the top-up injection mode. Solenoid-based Siberian snakes are being considered to maintain the polarization during acceleration in the booster, which was preliminarily verified using lattice-independent simulations. A pair of solenoid-based spin rotators were implemented around each IP in the e- collider ring, in an anti-symmetric manner, to realize longitudinal polarization direction at IPs while maintaining vertical polarization in the arc regions. Each solenoid section with a length of about 100 m, includes interleaved solenoid and quadrupole magnets for coupling compensation and optics matching, is added adjacent to the interaction region (IR). In the CDR lattice design the crossing angle is 16mrad, additional FODO cells with bending magnets are required to be inserted between IR and the solenoid section to compensate for the proper spin rotation angle. Simulations using the SLIM formalism implemented in Bmad shows above 80% equilibrium beam polarization at working beam energies with a fractional spin tune of 0.5. The polarization build-up time is much larger than the beam lifetime of 1-2 hours. Therefore, a larger than 50% time-averaged beam polarization in the top-up injection operation is possible if the polarization loss in the booster is well controlled. The implementation of the spin rotator leads to a moderate shrink of the dynamic aperture, which we believe can be recovered by dedicated optimization using more familities of sextupoles.

### 1.4.10 CEPC technology required and readiness

CEPC as circular electron positron Higgs factory collider composed mainly 20GeV S-band and C-band 20GeV injector linac, 100km booster and collider ring with MDI and SRF regions, it involves both conventional and advanced accelerator technologies, and some of them are listed below with the status of readiness.

The CEPC accelerator main key technologies required are listed below with prototype timelines:

1) CEPC 650MHz 800kW high efficiency klystron (77~80%) (at the end of 2021 complete the fabriation, finish test in 2022)

2) High precision booster dipole magnet (critical for booster operation) (Complete real size magnet model in 2022)

3) CEPC 650MHz SC accelerator system, including SC cavities and cryomules (Complete test cryomodule in 2022)

4) Collider dual aperture dipole magnets, dual aperture qudrupoles and sextupole magntes(Complete real size model in 2022)

5) Vacuum chamber system (Complete fabrication and costing test in 2022)

6) SC quadrupole magnets including cryostate (Complete short test model in 2022, dual aperture SC quadrupole in 2025)

7) MDI mechanic system (Remote vacuum connection be started in 2022)

8) Collimator (be started in 2022)

9) Linac components (Complete key components test in 2022, C-band linac be started in 2022)

10) Civil preliminary engineering design (Reference implementation design complete in 2022)

11) Plasma injector (Alternative injector technology, start electron acceleration test in 2022)

12) 18KW@4.5K cryoplant (Company) (to be ompleted in 2025)

13) SppC related high field superconducting magnets

More items are also under study and development, such as beam instrumentations, kickers, control system, alignment, mechanical supports, collimators, beam dump, etc.

### 1.4.11 CEPC environment impact

CEPC tunnel consists mainly two parts, the linac tunnel on the ground and the tunnel underground to host the booster and collider rings. The radiation doses both in linac and collider tunnel during operation have been estimated as shown in Figs.

8(a)(b) and safety measures will be taken to ensure the protection of the machine and environment. And underground water distribution quantity requirement will be provided during site selection and underground condition investigation.

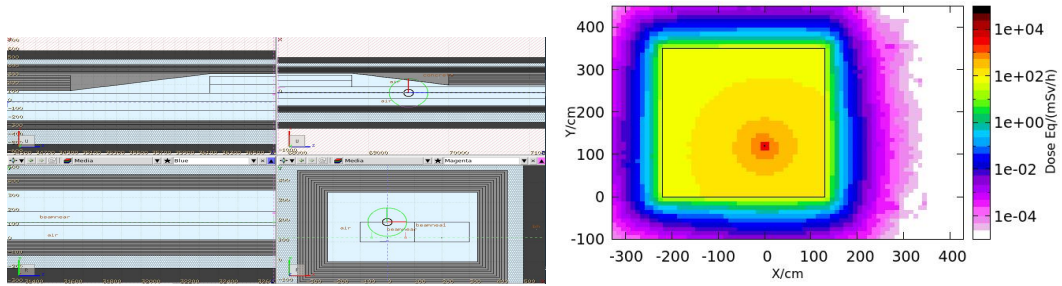

Fig. 8 (a) Dose equivalent simulation inside linac tunnel

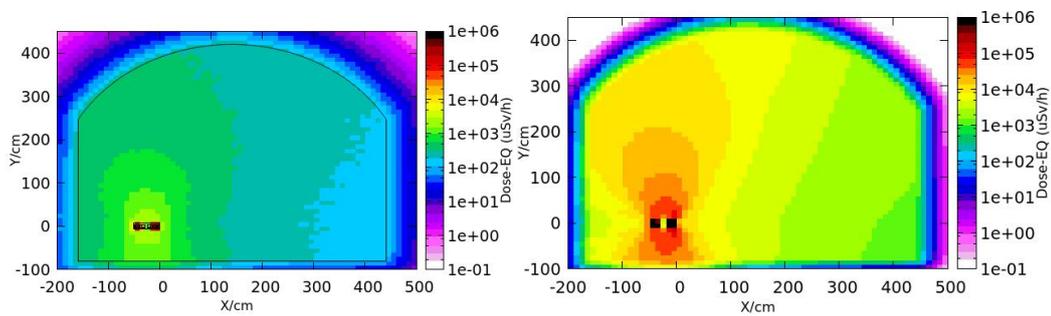

Fig. 8 (b) Dose inside tunnel: CEPC prompt radiation dose equivalent caused by random beam loss (left) and prompt radiation dose equivalent caused by synchrotron radiation (right)

## 2. Technology Requirements and R&D Status

### 2.1 Technology readiness Assessment

The CEPC key technologies in collide ring, booster and linac injector have been put to intensive R&D in TDR phase, which will be completed at the end of 2022, and after TDR, there will be several years EDR to make engineering design and mass production preparation before starting of the construction. Considering the progress and the proto type results of the key technology R&D, it is concluded that CEPC accelerator technologies will be in hand and ready for construction around 2026.

### 2.2 Required R&D

The detailed R&D and progresses are shown in the following subsections to support the technology readiness assessment conclusion.

#### 2.2.1 High efficiency klystron and RF source

The CEPC two beam synchrotron radiation power is more than 60 MW, it needs high efficiency RF source to minimize CEPC AC power consumption. Considering the klystron operation lifetime and power redundancy, a single 650MHz 800 kW klystron amplifier will drive two of the collider ring SC cavities through a magic tee and two rated circulators and loads. The CEPC high efficiency 650MHz klystron design goals are to set the efficiency to be above 80% and successful industrialization.

IHEP is developing 650MHz klystron with 800kW CW output power and 80% efficiency. To achieve this goal, a couple of klystron prototypes are being manufactured presently. The first prototype has been completely developed in 2020 with traditional bunching method with the efficiency reaching up to 62%, as shown in Fig. 9. The high efficiency klystron prototype has also been developed at the end of 2021 with the output power and efficiency to be 800kW and 75% respectively, as shown in Fig. 10. This prototype is being high power tested on the site of Platform of Advanced Proton Source (PAPS). Moreover, a multi-beam klystron prototype is being developed with designed efficacy of higher than 80%. The design schemes of high efficiency klystron with other methods are also in progress.

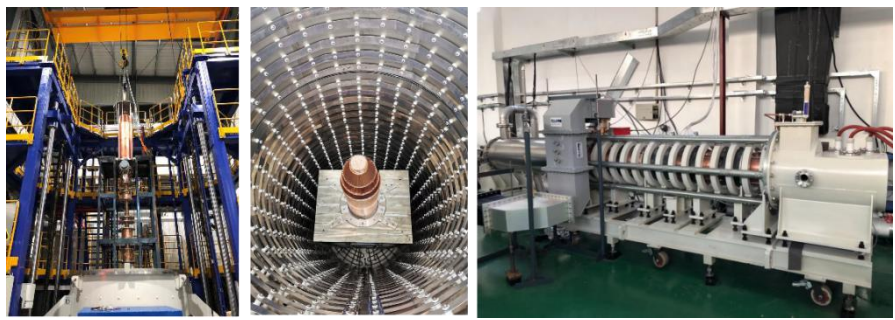

Fig. 9    CEPC 650MHz CW 800kW klystron (65% efficiency designed and 62% achieved)

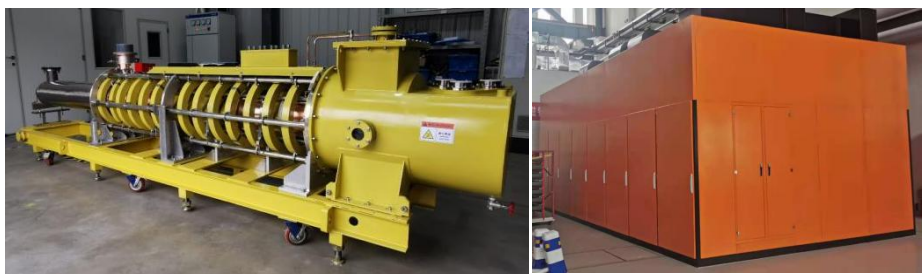

Fig. 10 CEPC 650MHz CW 800kW klystron and power source (77% efficiency designed and 75% expected in the test)

### 2.2.2 SRF technology

The CEPC SRF technical challenges that require R&D include: achieving the cavity gradient and high quality factor in the real cryomodule environment, robust and variable high power input couplers that are design compatible with cavity clean assembly and low heat load, efficient and economical damping of the HOM power with minimum dynamic cryogenic heat load, and fast RF ramp and control of the Booster.

Impressive test results are obtained on CEPC key SRF components and 650 MHz prototype cryomodule (Fig. 11 left) taking advantages of the new large SRF infrastructure (PAPS). CEPC design goal with world-leading high Q and high gradient 650 MHz and 1.3 GHz cavities (Fig. 11 right) have been recently achieved at IHEP with novel recipes, as shown in Figs. 12(a)(b)(c). In synergy with CEPC SRF R&D, large CW XFEL projects in China, such as SHINE (Shanghai HIgh repetition rate XFEL aNd Extreme light facility) in Shanghai etc., will need a total of one thousand high Q 1.3 GHz 9-cell TESLA cavities and their cryomodules in next five years, while IHEP is playing a leading role in the key technology development in this national SRF context. In parallel with design and key R&D, extensive development of SRF personnel, infrastructure and industrialization is essential for the successful realization of CEPC. Meanwhile, IHEP will maintain and extend CEPC SRF collaborations with international laboratories with strong expertise.

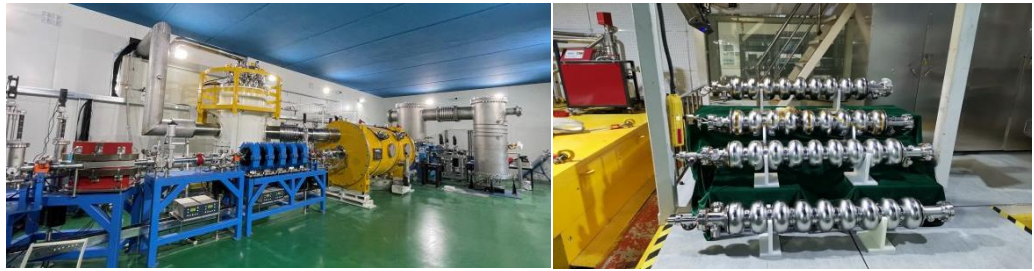

Fig. 11 CEPC 650 MHz test cryomodule with beam and 1.3 GHz 9-cell cavities

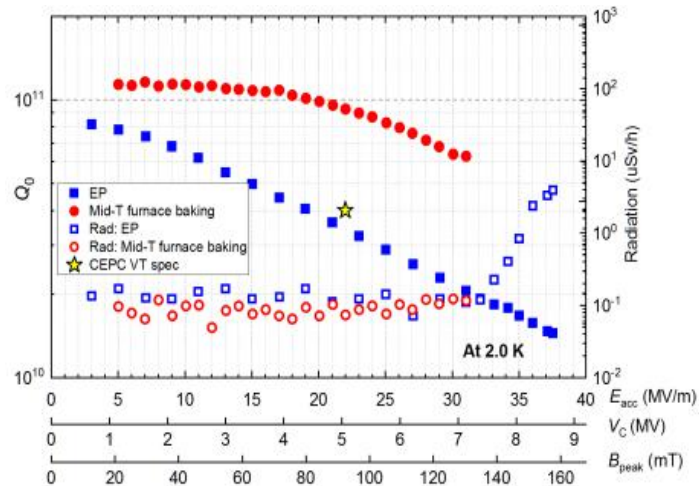

(a)

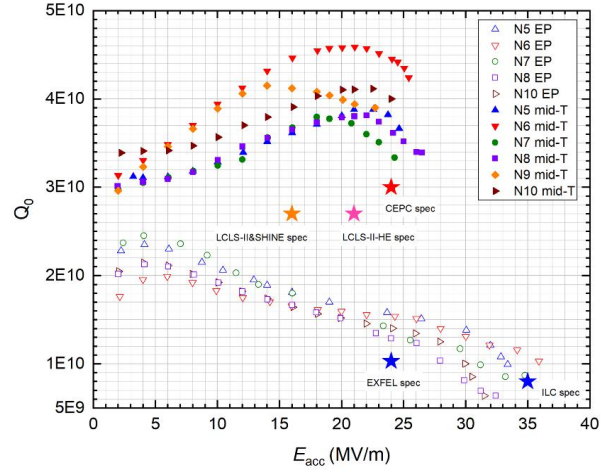

(b)

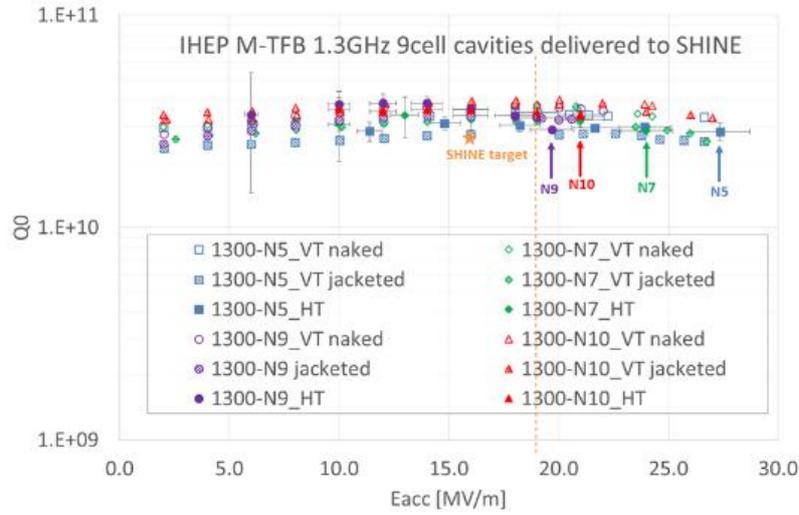

(c)

Fig. 12 CEPC 650 MHz single cell vertical test (a), 1.3 GHz 9-cell cavity vertical test (b) and horizontal test (c) performances

### 2.2.3 Cryogenic system description

R & D of the CEPC cryogenic system includes the superconducting radio frequency (SRF) cavity cryogenic system, the superconducting (SC) magnet cryogenic system, and several key technologies (high performance JT heat exchanger design/cryo-circulating liquid helium pump/low heat leak multi-channel transfer line research etc.).

The CEPC superconducting cavity cryogenic system side will be equipped four individual refrigerators with the capacity of 18kW@4.5K for each station. The cooling temperature of SRF cavity is 2K. For the Booster ring, there are 12 cryomodules for 1.3GHz 9-cell cavities. And for the collider ring, there are 56 cryomodules for 650MHz 2-cell cavities in the Higgs 50MW mode. In the previous conceptional design rate (CDR), the cooling scheme is parallel for the SRF cryomodules. Each cryomodule has a separated valve box in the gallery tunnel. Later in the technical design rate (TDR), to improve the cavity performances while reducing cost, the cooling scheme of both booster and collider cryomodules has been modified, from parallel to series. When compared the heat load of the two cooling schemes, it's found that the modified scheme can reduce 4kW heat load, and the length of multi-transfer lines can also be shortened largely. Furthermore, the design of both cryomodules has been reviewed and improved based on the growing experience of the IHEP cryogenics group. Prototypes for booster and collider have been built in collaboration with the domestic qualified companies and has been tested in the Platform of Advanced Photon Source (PAPS) infrastructure with the aim of further improvements. As for the large helium refrigerator market research, it's encouraged to adopt the domestic plant producer. The 18kW@4.5K large helium refrigerator is the new generation product for FULL CRYO company, which is expected to be fully prepared for the CEPC project.

The CEPC superconducting magnet cryogenic system has 4 interaction region (IR) magnets working 4K in the MDI equipped with 2 individual refrigerators with the capacity of 2.5kW@4.5K, and 2 detectors adopts thermosiphon cooling method with 2 individual refrigerators with the capacity of 1.5kW@4.5K. The choice of the use of large refrigerators for operational stability and cost was appreciated compared with the small GM refrigerators scheme. The design of 32 IR sextupole magnets are updated through reducing the magnet bore and pole field, so the cooling scheme has been changed from superconducting to normal conducting.

The CEPC cryogenic system has several key technologies, includes high performance JT heat exchanger design/cryo-circulating liquid helium pump/low heat leak multi-channel transfer line research/virtual system establishment and automatic control strategy research. **a)** the JT heat exchanger platform R&D and test has been successfully carried out, and higher flow rates of 2K heat exchangers (10g/s → 50g/s → 120g/s requirement for CEPC SRF cryogenic system) have been developing. **b)** the cryo-circulating liquid helium pump is the key equipment for the quadrupole iron with narrow cooling channels in CEPC collision zone, when adopting supercritical helium forced-flow cooling or superfluid helium cooling method. **c)** low heat leak multi-channel transfer line research platform has been established and researched, and 0.15W/m heat loss transfer lines needs to be developed for CEPC cryogenic system. It's expected to do the experiment test in the end of this year. **d)** the Model prediction method (MPC) as a control algorithm method has been successfully tested in PAPS BTCM cryostat, and neural network construction inversion model has been developed, which laid a good foundation for the CEPC cryogenic system R&D.

All in all, further TDR design is going on. Every member needs to work together to greatly push forward the work, and complete the TDR design on time. The CEPC cryogenic system for booster and collider ring SRF accelerators is shown in Fig. 13.

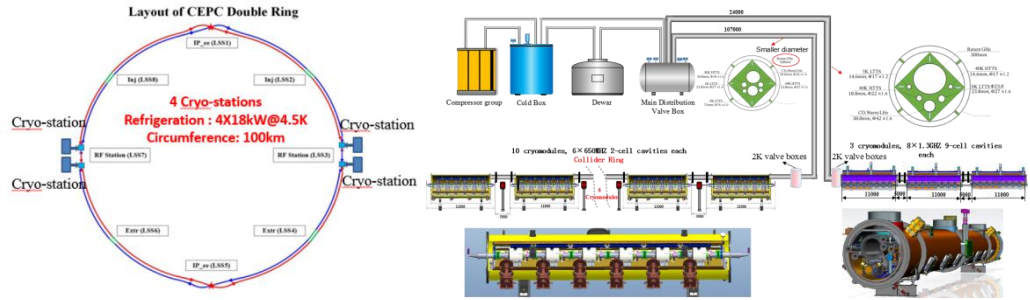

Fig. 13  CEPC cryogenic system for booster and collider rings' SRF accelerators

### 2.2.4 CEPC collider magnets

The CEPC collider ring is a double ring with a circumference of 100 km. There are 2466 dipoles, 3052 quadrupoles, 948 sextupoles and 2904 correctors in the CEPC collider ring. Theses conventional magnets occupy over 80% of collider ring. Therefore the cost and power consumption are two of the most important issues. Several aspects are considered, including the field cross talk effect, the synchrotron radiation, energy sawtooth effect and the mass production process. Most of the dipoles and quadrupoles are designed as dual aperture magnets which can save about 50% power. The aluminum coils are used as excitation coils. A short prototype magnet of dual aperture dipole with sextupole component was designed and measured as shown in Fig. 14(left). The prototype of long dipole magnet is in production. Special design is adopted to solve the cross-talk effect between two apertures of dual aperture quadrupole as shown in Fig. 14(middle). The short dual aperture quadrupole prototype will be measured later as shown in Fig. 14(right).

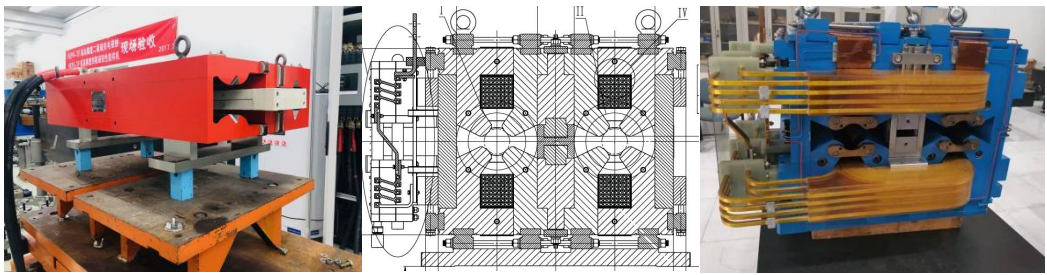

Fig. 14 CEPC dual aperture dipole(left) and dual aperture quadrupoles (right)

### 2.2.5 CEPC booster magnets

The circumferences of the booster is about 100 km, which has 14866 dipoles, 3848 quadrupoles, and 350 correctors. The gap of the dipole magnets is 63 mm, the most of them are 4.7 m long, the others are 2.4 m and 1.7 m long. The field will change from 63 Gauss to 564 Gauss during acceleration. The field errors in good field region are required to be less than 1E-3. Due to very low field level, the cores are composed of stacks of three 0.3 mm thick oriented silicon steel laminations spaced by 1 mm thick aluminum laminations. Since magnetic force on the poles is very small, the return yoke of the core can be made as thin as possible. In the pole areas of the laminations, some holes will be stamped to further reduce the weight of the cores as well as to increase the field in the laminations. All above considerations can improve the performance of the iron core and considerably reduces the weight and the cost. Also for economic reasons, the excitation coils that have two turns are made from pure aluminum bars with the cross section 30×40 mm² without water cooling.

The bore diameter of the quadrupole magnets is 63 mm, the magnetic length of the 2/3 quadrupole magnets is 2 m, the length of the others is 1 m. The max. quadrupole field is 18.05 T/m whereas he min. quadrupole field is 1/12 of the max. field. For cost reduction, the hollow aluminium conductors instead of copper conductors are selected to wind the coils. The iron cores are made of 0.5 mm thick low carbon silicon steel laminations. The magnet will be assembled from four identical quadrants, and can also be split into two halves for installation of the vacuum chamber.

For the new lattice of the CEPC booster, there are not sextupole magnets.

The gap of the correctors is 63 mm, the max. field is 200 Gs, the field errors in good field region is required to be less than 1E-3. To meet the field quality requirements, the correctors have H-type structure cores so the pole surfaces can be shimmed to optimize the field. The cores are stacked from 0.5 mm thick laminations. The racetrack shaped coils are wound from solid copper conductor. Each coil has 24 turns, which are formed from 4 layers; no water cooling is required.

Two types of short dipoles for the booster have been fabricated and tested as shown in Fig. 15, both could satisfy the design goal with 20GeV injection energy, and real size prototype is under fabrication.

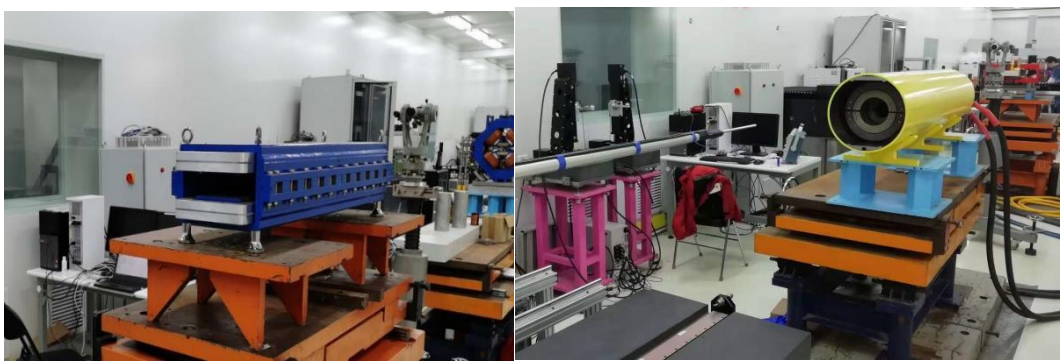

Fig. 15 Two types of CEPC booster magnets, with iron core (left) and without iron core (right)

## 2.2.6 CEPC interaction region superconducting magnets

Compact high gradient final focus superconducting quadrupole doublet magnet (QD0 and QF1) are required on both sides of the collision points of CEPC collider ring. QD0 and QF1 are double aperture quadrupoles and are operated fully inside the field of the Detector solenoid with a central field of 2~3T. To minimize the effect of the longitudinal detector solenoid field on the accelerator beam, anti-solenoids before QD0, outside QD0 and QF1 are needed. The total integral longitudinal field generated by the detector solenoid and anti-solenoid coils is zero. It is also required that the total solenoid field inside the QD0 and QF1 magnet should be close to zero. The superconducting QD0, QF1, and anti-solenoid coils are in the same cryostat, which makes up a combined function magnet. In the TDR R&D phase, superconducting prototype magnets for the CEPC interaction region will be developed in three consecutive steps: 1) Double aperture superconducting quadrupole prototype magnet QD0 (136T/m, 2m long). 2) Short combined function superconducting prototype magnet including QD0 and the anti-solenoid. 3) Long combined function superconducting prototype magnet including QD0, QF1 (110T/m, 1.48 long) and anti-solenoid. The CEPC 0.5m superconducting QD0 test model with field gradient of 136T/m is under development, and the first collared quadrupole coil has been fabricated as shown in Fig. 16.

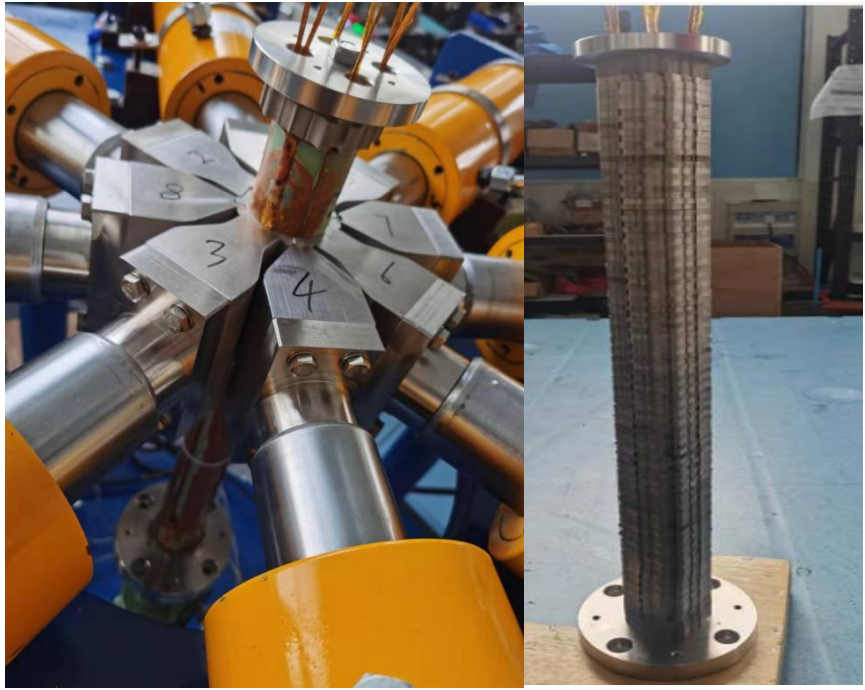

Fig. 16 CEPC first collared quadrupole coil for 0.5m QD0 test model

## 2.2.7 CEPC vacuum system

CEPC vacuum system requirements for booster and collider rings are shown in Table 11.

Table 11: CEPC vacuum system requirements

| Zone | Material | Vacuum/ Torr | Cross Section |
|------|----------|--------------|---------------|
| Booster | Extruded aluminum 6061 | $3 \times 10^{-8}$ | φ56 |
| Electron Ring | Extruded aluminum 6061 | $3 \times 10^{-9}$ | 56×75 to φ56 |
| Positron Ring | Extruded copper, NEG film | | 56×75 to φ56 |
| MDI | Copper/tungsten alloy, NEG film | $3 \times 10^{-10}$ | φ20 |

R & D of the CEPC vacuum system include the vacuum chambers, RF shielding Bellows and NEG coating inside the inner surface of the copper vacuum chambers in the positron ring.

The Collider will have an aluminum chamber for the electron beam and a copper chamber for the positron beam. The aluminum chamber has a beam channel, a cooling water channel, and thick lead shielding blocks covering the outside. Between the vacuum chambers several pumping ports used to install ion pump, bulk NEG pump, and gauges. The copper chamber has a beam channel and a cooling water channel, and NEG coating will be used. In the R&D program the prototypes of both copper and aluminum vacuum chambers will be fabricated and tested. The ultimate pressure of the vacuum chambers is less than $3 \times 10^{-10}$ Torr；The thermal outgassing rate is less than $1 \times 10^{-12}$ Torr•L/s/cm$^2$.

The NEG coating is a titanium, zirconium, vanadium alloy film, deposited on the inner surface of the chamber through magetron DC sputtering. R&D is required so the sputtering process for NEG film deposition is optimized to avoid instability and lack of reproducibility. These problems can significantly change the gas sorption and surface properties (e.g. secondary electron yield, ion-induced gas desorption). During tests of the coating, all related parameters (plasma gas pressure, substrate temperature, plasma current, and magnetic field value) will be recorded and suitably adjusted to ensure stability of the deposition process. After coating, the chambers will be activated at different temperature to test its pumping speed of $H_2$ and CO. The design specification with pumping speed of 0.5 L/s/cm$^2$($H_2$) for the coating vacuum chamber will be achieved.

To eliminate the quadrupolar wakes, elliptical vacuum chamber in the collider ring in CDR will be replaced by circular chambers with dimeter of 56 mm in TDR. The prototypes of copper & aluminum vacuum chambers with a length of 6 m have been fabricated and tested, which meet the engineering requirements, as shown in Fig. 17(a).

The primary function of the RF shielding bellows is to allow for thermal expansion of the chambers and for lateral, longitudinal and angular offsets due to tolerances and alignment, while providing a uniform chamber cross section to reduce the impedance seen by the beam. The usual RF-shield has many narrow Be-Cu fingers that slide along the inside of the beam passage as the bellows is compressed. The design specification of the contact pressure for RF contact fingers is 125±25g/finger. The prototypes of RF shielding bellows have been fabricated. The key components experiment such as spring fingers and contact fingers have been carried out. Contact force is uniformly from different fingers and meets the target of 125±25g, as shown in Fig. 17(b).

Surface treatments of vacuum chamber is undertaking. A 2m long vacuum pipe have been NEG coated to explore the coating parameter at geometrical shape of 56×75, and vacuum reached $2.5 \times 10^{-10}$ Torr after 200℃/24h activation. The 6m long vacuum chambers will be NEG coated by moving the solenoid with a vertical coating setup. The setup of NEG coating which has ability to coat 6 meters long pipe by moves solenoid is being built for vacuum pipes at location of IHEP Dongguan branch in Guangdong province, as shown in Fig. 18. Tungsten alloy will be used to fabricate the fork vacuum chamber of MDI.

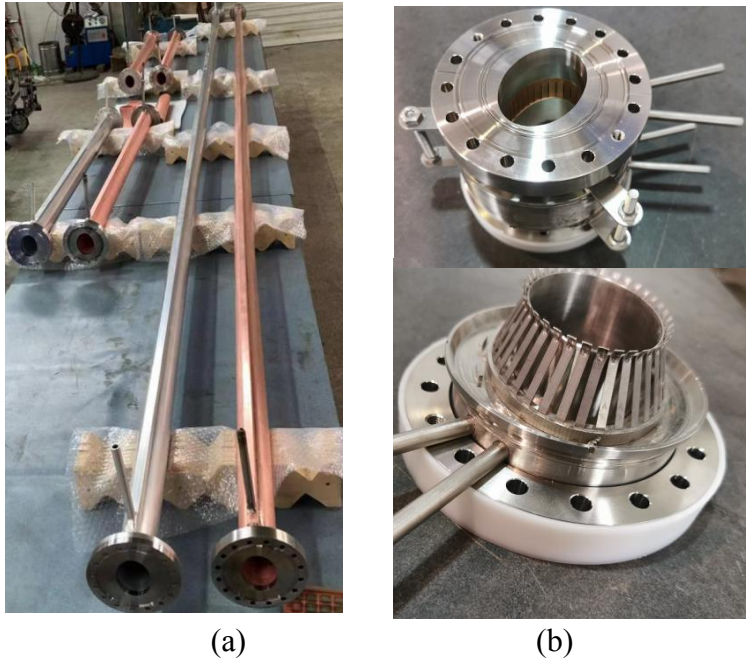

<div align="center">(a)           (b)</div>

Fig. 17 CEPC copper/aluminium vacuum pipes of 6m long(a) and RF shielding bellows (b)

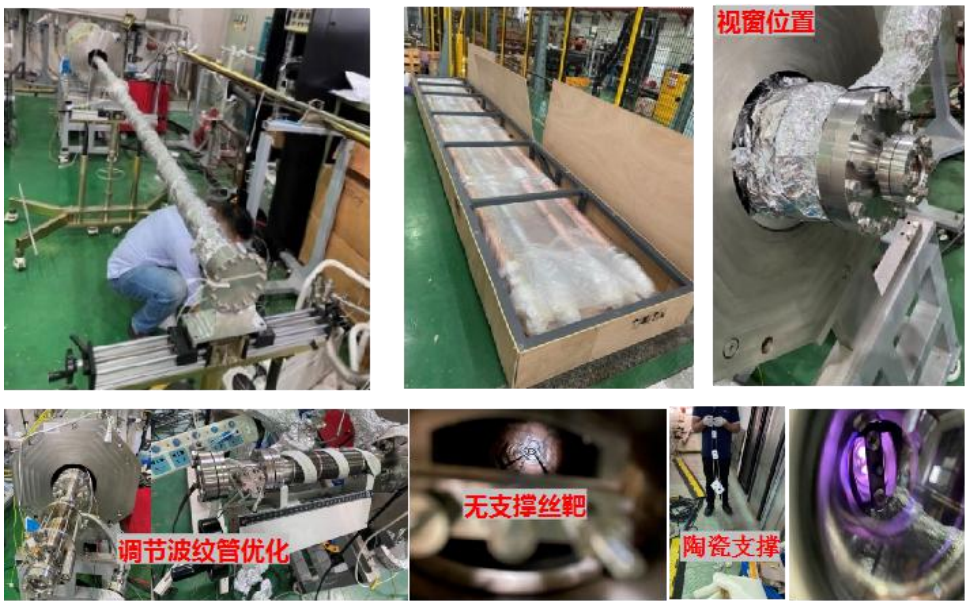

Fig. 18 CEPC vacuum chamber NEG coating system

### 2.2.8 Electrostatic-magnetic separator

In the RF region of CEPC, the RF cavities are shared by the two rings. A set of electrostatic separators combined with dipole magnet (Electrostatic-Magnetic Separator) near by the RF cavities are used to avoid bending of incoming beam and deflect the outgoing beam in H mode. The Electrostatic-Magnetic Separator is a device consisting of perpendicular electric and magnetic fields. The key issues of the separator's design are: homogeneous field design; to maintain E/B ratio in fringe field region; reduce the impedance and loss factor of the separator; mechanical structure design.

The electrostatic separator unit consists of a pair of pure Titanium electrodes — each 4 m long and 180 mm wide — mounted in an UHV tank of about 380 mm inner diameter. The magnet yoke is H-type, the center magnetic field is 66.7 Gauss. Within the patch of 6cm*11cm, the uniformity of the field integrals reaches ±2E-04.

We have completed the design of Electrostatic-Magnetic Separator both electrical and mechanical. It is including: the simulation analysis of the parameters of the electrostatic separator; the simulation analysis of the installation error of electrostatic separator; the impedance optimization for Electrostatic-Magnetic Separator; the thermal analysis of the electrode plate; the design of the cooling system of the electrodes; the design of the vacuum system; the mechanical design of the separator and the dipole magnet.

The prototype of electrostatic separator and dipole magnet have been fabricated as shown in Fig. 19. And the overall installation of the separator was completed and test had been done. The test result showed that the vacuum reached the target ($.\leqslant 1.0 \times 10^{-10}$Torr). But the high voltage test failed due to the arc occurred at the air side of feedthrough. So, after the improvement by feedthrough, the separator test will continue.

The prototype of magnet has been fabricated, too. When the test platform is set up, the magnets will be tested.

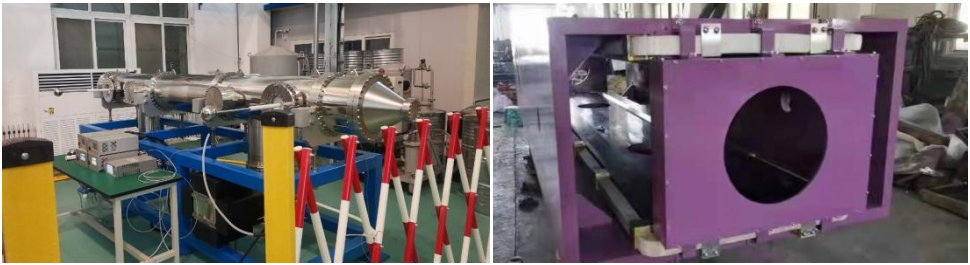

Fig. 19 The prototypes of CEPC electrostatic separator and dipole magnet

### 2.2.9 Linac injector key technologies

The CEPC linac injector is a normal conducting S-band (2860MHz) & C-band (5720MHz) linac. It provides electron and positron beams at an energy of 20 GeV to the booster. The linac injection key technology including electron source, positron source, SHBs, buncher, S-band RF system, C-band RF system and the damping ring 5 cell cavity.

A conventional thermionic electron gun is chosen. Positron source choose a fixed target. For the flux concentrator produces a magnetic field with a sharp rise over less than 5 mm, a prototype is made and have finished the high power test.

The bunching system is consists of the SHBs and buncher . The first subharmonic buncher (SHB1) operating at 158.89 MHz (18th subharmonic), the second subharmonic buncher (SHB2) operating at 476.67 MHz (6th subharmonic), and a constant-impedance travelling-wave buncher operating in $2\pi/3$ mode at 2860 MHz. All the three elements have been simulated and designed.

For the main linac, the S-band constant gradient accelerating structure with $2\pi/3$

and pulse compressor have designed and manufactured. The S-band accelerating structure has finished high power test at the high power test bench. An S-band spherical cavity pulse compressor has been manufactured now. A C-band constant gradient accelerating structure with $3\pi/4$ mode has been manufactured and tuned. The CEPC linac injector key components are fabricated as shown in Figs. 20 and 21.

There is a damping ring at 1.1 GeV in the linac. It includes two 5cell RF cavity. RF design and mechanical design of cavity, coupler and tuners have finished.

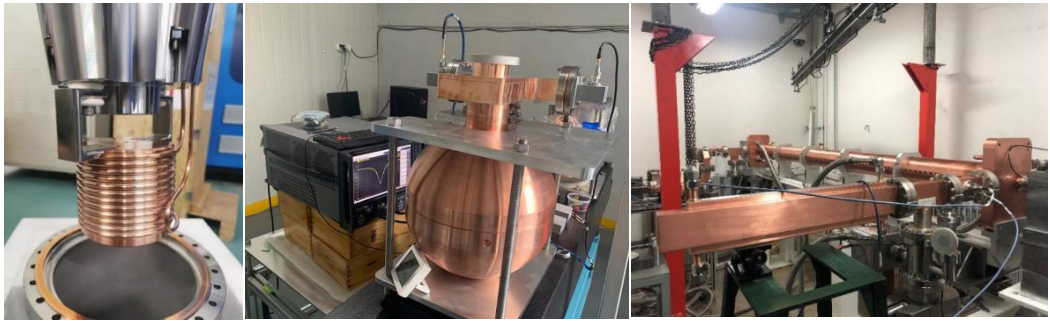

Fig. 20 CEPC linac injector positron source flux concentrator, S-band SLED pulse compressor and high gradient S-band accelerating structures

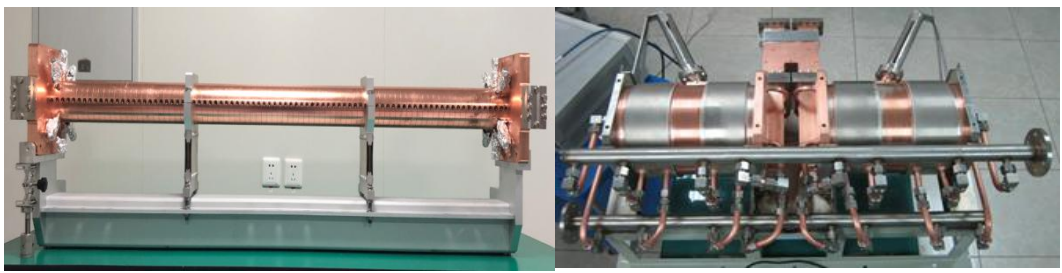

Fig. 21 IHEP C-band accelerating structure and C-band SLED

## 2.2.10 Mechanical supports, RVC and collimators

Mechanics system mainly focus on the magnet support design, collimator design, MDI RVC (remote vacuum connector) design and the device layout design.

There are over 10 thousand of dipole magnet yokes and thousands of quadrupoles, sextupoles and other magnets in collider ring, and also plenty of magnets in booster, Linac and transport lines. All of the magnets need supports. Besides, the high luminosity needs the high stability and alignment accuracy of magnets, for example, 100 um alignment tolerance and 4 nm vibration amplitude at MDI regions. The supports should have simple structure with good flexibility and kinetic performance for installation and alignment, high stability for vibration resistance and low cost.

The collimators are for decreasing the background and facility protection. The synchrotron radiation power is about 9.3 kW at 30 MW, Higgs mode, and will scale up at 50 MW. If beam failure happens, only 0.3 bunch can make the vacuum chamber melting. New methods should be studied for the heat dissipation and resistance for thermal shock.

The RVC is the remote vacuum connector at MDI region, for the vacuum connection of the detector chamber and the accelerator chamber. The long cantilever and limited operation space make the design of RVC much more difficult than the existing facilities like SKEKB or BEPCII. We have designed several structures and now are focusing on the improved inflatable sealing. The limited dimension and high vacuum leak rate are the main difficulties.

There are plenty of devices for CEPC, the device layout is very important, especially at MDI regions. The purpose is to optimize the space usage that all the devices have the suitable space for operation and maintenance. It is also a reference for device design and installation preparations.

The CEPC mechanical supports, collimator and RVC in MDI designs are shown in Fig. 22.

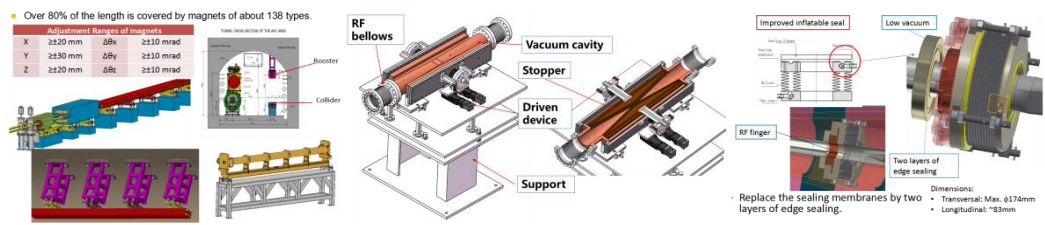

Fig. 22 CEPC mechanical supports, collimator and RVC in MDI

## 2.2.11 Beam dumps, radiation shielding and machine protection

A beam dump with a set of dilution kickers is designed for the collider's normal running. It has three layers: absorb core, iron, and concrete shell. The maximum temperature rises in the dump are simulated using FLUKA for Z-pole energy, $W^+W^-$ threshold, Higgs, and ttbar operations. This dump system can be installed around four linear segments of the collider. The linac dumps are also designed. Based on CEPC CDR parameters, we estimate the dump dimensions that meet the upper limit, 5.5mSv, out of these dumps.

Another key issue is synchrotron radiation shielding. To evaluate its impact on the magnet coils and tunnel environment, FLUKA simulations were performed for Z pole, WW threshold, Higgs, and ttbar operations. It is shown that the lead shielding scheme can reduce the absorbed dose of the coils. The dose distributions and residual nuclei productions in the circular and linac tunnel are also obtained.

Machine protection is a key issue for CEPC, experiences from super KEKB and LHC, and other high-energy colliders/storage rings are necessary.

## 2.2.12 CEPC control system

The CEPC control system should control and monitor all the equipment with friendly OPIs, robust and efficient communications, strong beam-tuning tools and rich application tools to achieve the desired beam performance.

The CEPC control system consists of global control system, local control system and the third party system integration. The global control system includes Timing system, MPS, control network, computers and servers, management of information and database, etc. Local control system is composed of power supply control, vacuum control, RF control, Injection/Ejection control, temperature monitoring and so on. The party system integration refers to interface with conventional facility, experimental detector, control of commercial equipment and so on.

To build up so large a control system, the more commercial industrial products and technique are adopted, the better quality the whole project will be. With the evolution of electronic techniques, hardware prices decrease rapidly, while with much better performance. So, purchase and final mass-production should be made as later as possible. On the other hand, technical studies and interfaces between different systems should be made as earlier as possible to ease the system development, integration and commissioning. A full-scale prototype system should be set up first for development and function tests.

### 2.2.13 CEPC beam instrumentations

For CEPC beam instrumentation a full spectrum R&D have been carried out, such as, BPM electronics, beam position monitor fabrication, longitudinal feedback system, transverse feedback system, synchrotron radiation monitor, BI at the interaction point, bunch current monitor, and beam loss monitor, etc. As shown in Fig. 23. Among all CEPC beam instrumentation, the number of beam positions is the largest, about 5,000 sets. So in the TDR stage, we focuses on the development of beam position monitor (BPM) related technologies, including BPM electronics and pick-ups.

Beam position monitor is an indispensable key system of modern accelerator. It can not only measure beam position and monitor beam orbit, but also be used to calculate other important physical parameters. "High-speed, high-resolution AD sampling technology" and digital signal processing methods such as modulation, demodulation, filtering. These technologies and methods can also be directly applied to the research fields of military communication, radar detection, aerospace and medical imaging. Over the past few years, with the joint efforts of more than a dozen backbone scientific researchers and postgraduates, the self-developed digital BPM electronics has been tested in laboratory (the electronics resolution can reach 10 nanometers) and actual beam flow. The test results show that the main performance indicators have reached or surpassed those of similar foreign commercial products. In 2019, self-developed digital BPM Electronics was first successfully used in BEPCII linac accelerator. In 2020, self-developed digital BPM electronics was also successfully applied in BEPCII storage ring and had been decided to be applied in the HEPS project under construction. Until now, more than 100 sets BPM electronics is operating in BEPCII linac and storage ring. In addition, based on the research and development of digital BPM technology, the R&D team has developed Bunch by Bunch beam measurement electronics. High-speed ADC and fast data processing algorithms are used for fast measurement and analysis of beam parameters to study the physical quantity that changes rapidly with the beam.

More efforts was paid on the feed through which is a key component of BPM pick-ups. After studying feed through, two versions of feed-through have been made with the help of the CIPC Member Company. Many tests were done to check the characteristic of feed through, including the mechanical properties, high-frequency characteristics and vacuum performance. More studies will be done on the consistency of batch products.

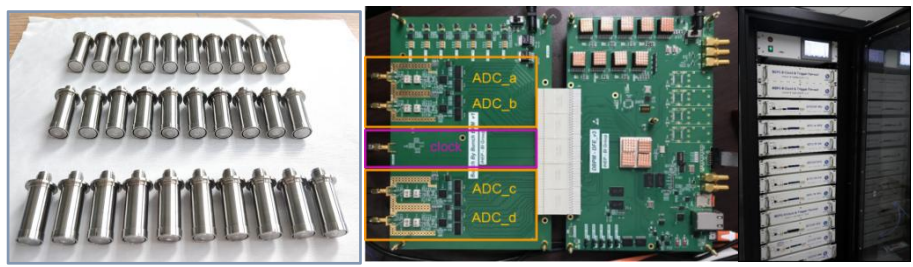

Fig. 23 CEPC beam instrumentation feed through and electronics

### 2.2.14 Septum, kickers, and power supply

The CEPC consists 9 injection and extraction sub-systems, from the linac to the damping ring, the booster ring and the collider rings. In order to realize the beam injection and extraction under different beam energy and different beam filling pattern for each rings with different machine acceptance, the parameter requirements of each subsystem are very different, so different types of hardware design must be adopted, including the septas, kickers and pulsed power supplies.

Lambertson septa is the first choice for CEPC because it is more reliable than other types of septas with typical septum wall width from 2mm to 10mm.

Which type of kicker to been chosen is determined by the injection scheme, the waveform of pulse and required kick strength. The waveform of kicker pulse is determined by injection scheme, filling pattern and revolution period. For the damping ring, a slotted-pipe kicker driven by a pulser with 250ns half-sine are adopted. For the booster low energy injection, a set of strip-line kickers and 50ns half-sine pulsers are used for bunch-by-bunch injection. In W and Z mode, the beams are extracted train by train from the booster then injected into the collider rings, so a trapezoid delay-line kicker system with adjustable pulse width range from 440ns to 2420ns is appropriate. In Higgs mode, due to smaller dynamic aperture of collider rings, a kind of on-axis swap-out injection scheme is adopted. Here, beam accumulation is realized in the booster at full energy. In such injection process, bunches are injection and extraction one-by-one. Accordingly, the kicker with 1360ns half-sine pulse bottom width is qualified for such operation mode. According to the pulse power technology developing trend, the solid-state switch is preferred for the future project.

The injection and extraction hardware R&D activities are carried out. In fact, one team is in charge of both HEPS (High Energy Photon Source) and CEPC inj. & ext. system In IHEP. A part of hardware R&D for 2 projects are overlapping including thin septum Lambertson magnets and fast kicker system. Some experience of R&D could be shared. For example, a half in vacuum Lambertson magnet with 2mm septum wall and an in air Lambertson magnet with 3.5mm septum wall, strip-line kicker system with 10ns pulse width and slotted pipe kicker system with 200~300ns pulse width for HEPS can meet the requirements of CEPC. Fig. 24 shows the slotted pipe kicker prototype.

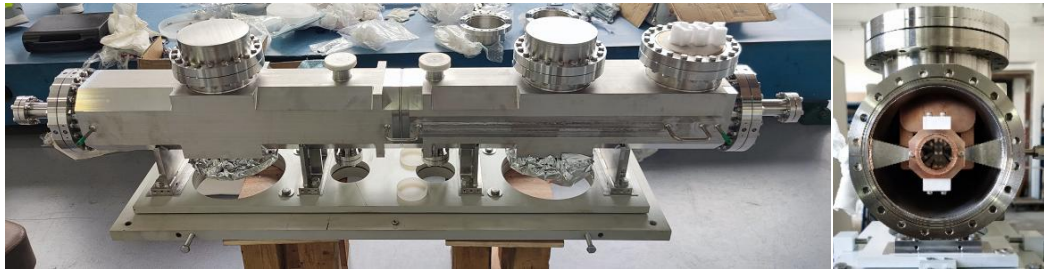

Fig. 24 CEPC slotted pipe kicker prototype

### 2.2.15 Alignment and installation preparations

CEPC alignment work mainly includes global datum construction, control network building and measurement, component fiducialization, pre-alignment, component installation and alignment and smooth alignment. The alignment accuracy for the primary magnets in the arc areas is 0.1mm in transversal vertical and longitudinal, and 0.1mrad for pitch yaw and roll angles. To meet these requirements, it needs to establish high accuracy global datums and apply special technologies for CEPC alignment.

The diameter of CEPC collider ring is about 32km, for such big range alignment, it needs three global datums to realize the global position control. The first is the reference ellipsoid, it is the base to establish the projection plane, Quasi-Geoid model and Vertical deflection model. In China the CGCS2000 reference ellipsoid can be used. The second is the Quasi-Geoid model, it is the datum for the height coordinate calculation. To improve the height position accuracy control, the precision of the Quasi-Geoid model should be better than 5mm. The third is the Vertical deflection model, it is the datum for the vertical axis of the measuring station coordinate system directional in the CEPC coordinate system. The precision of the Vertical deflection model should be better than 1" in North-South and East-West respectively.

To provide the coordinate reference system and control error accumulation, a three-levels control network was designed, it includes the surface control network, the backbone control network and the tunnel control network. The tunnel control network is the reference for component installation and alignment, its relative position precision between adjacent points is about 0.07mm. Component fiducialization is to relate the beam center with its fiducials, generally the measurement is based on its mechanical surface and the accuracy is about 0.05mm-015mm. To improve the magnet fiducialization accuracy, the technique base on magnetic field measurement can be adopted. Pre-alignment is to align the components on the girder in advance, then the cell can be installed and adjusted as one unit. Pre-alignment method can get higher relative position accuracy on one girder and can save the installation time. Component installation and alignment will use the tunnel control network as a position reference and use laser tracker to carry out the measurement.

Component installation will be divided into two phases; each phase is half a ring. In each phase, the installation will be divided into four sections to be carried out in parallel, and the length of each section is 12.5km. There will be a transport shaft in the middle and ends of each section respectively. For each section, a segments parallel installation strategy will be adopted to improve the installation efficiency.

A logistic and installation management system based on installation schedule, supply chain and logistics is under studying. According to the simulation, around the ring 8 warehouse areas and 8 assembly and testing areas will be built, and 9 warehouses should be built in each warehouse area, the area of each warehouse is 3000-4000m2. For the Linac, 1 warehouse area and 1 A&T area will be built. The transport of cryomodule and collider ring long dipole needs special attention. The long dipole will be divided into 5-6m length parts for transport and installation. Special facilities should be made to fix and protect the cryomodule. The transport parameter of the cryomodule includes the maximum dimension 11×Ø1.5 m, maximum weight 15T, maximum tilt ±15°, maximum shock 1.5g, roads and transport vehicles should meet these requirements. Specialized vehicles will be used for components transport and installation in the tunnel, the vehicles need further R&D.

In order to increase the alignment and installation spead, a new alignment instrument has been invented by IHEP, i.e. three functions, 1) photogrammetry, 2) distance, 3) angle measurement, all three in one instrument as shown in Fig. 25.

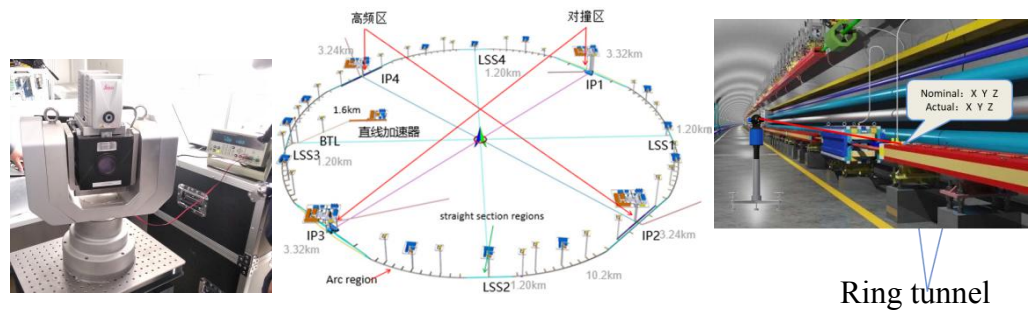

Fig.25 CEPC three functions in one instrument for accurate and efficient alignment

## 2.2.16 CEPC BIM design

CEPC has started Building Information Modelin (BIM) design which includes tunnel system, components, cabling etc. to help the engineering design, fabrication and installation, as shown in Fig. 26. Three BIM design videos are given in Refs. 14-16.

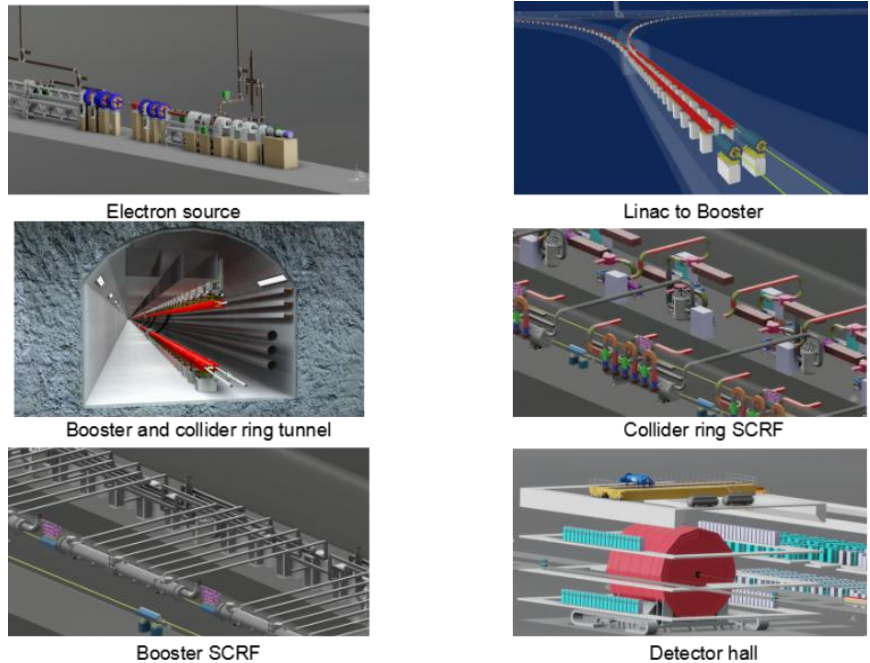

Fig. 26 CEPC BIM design

## 2.2.17 CEPC plasma injector (alternative technology)

The CEPC Plasma Injector (CPI) was proposed as a backup solution to the low field dipole problem of the booster. In CEPC conceptual design report, the electron/positron energy is 10 GeV at the linac exit. The particles need to be accelerated from 10 GeV to 45.5 GeV in CPI, and then injected to the booster. Although plasma wakefield acceleration (PWFA) has been studied more than 40 years and got tremendous progress recently, it still needs to address many issues if we try to use it in a real project.

First of all, simulation studies on PWFA and error tolerance analysis for electrons are performed. Two set of beam/plasma parameters are given for high transformer ratio (TR~4, 10→45.5 GeV) condition and normal transformer ratio (TR~1.5, 10→25 GeV) condition, respectively. For HTR condition, extra damping mechanism must be introduced to suppress the severe hosing instability, while the NTR mode seems more stable and realistic.

Another key issue for CPI is stable and high efficiency PWFA for positrons. The so-called "bubble regime" which is widely employed in electron PWFA, is not fit for the positron acceleration. After theoretical and simulation analysis, we find that an asymmetric electron driver can excite a perfect wakefield for positron acceleration in a hollow plasma channel with a high efficiency (≥30%). These results have been published in **PRL**, **127** 174801 2021, as editor suggestion, and attracted wide attention.

Besides the above on-paper work, we also made some important progresses in the experiments. We achieved "perfect" external injection with almost 100% efficiency in 2019, and use a plasma dechiper to decrease the beam energy spread from 1% to 0.1% in 2020. Both of these results are crucial for "connecting" a conventional accelerator and a plasma accelerator.

For the next step, we schedule to finish the start-to-end simulation including complete error tolerance analysis in 2022. On the other hand, although it's quite difficult to find a proper beamline to demonstrate the CPI's feasibility, we will try our best to experimentally prove the reliability of our simulation-based design.

## 2.2.18 High field superconducting magnets for SPPC

All the superconducting magnets used in existing accelerators are based on NbTi technology. These magnets work at significantly lower field than the SppC 12~24 T required by SppC. The upcoming 11 T dipole magnets for HL-LHC project are state-of-the-art superconducting magnets for accelerators.

SppC demands advanced or new type of superconducting materials with low cost and capable of applying in the high fields. Since 2008, iron-based superconductors (IBS) have been discovered and attracted wide interest for both basic research and practical applications. It has high upper critical field beyond 100 T, strong current carrying capacity and lower anisotropy. In 2016, the Institute of Electrical Engineering, Chinese Academy of Sciences (IEE-CAS) manufactured the world's 1st 100-m long 7-filamentary Sr122 IBS tape with critical current of $1 \times 10^4$ A/cm$^2$ at 10 T successfully, which makes the possibility of fabricating real IBS coils. In 2018, IHEP and IEE fabricated the IBS solenoid coil and tested at 24 T successfully. In 2018 and 2019, IHEP fabricated the IBS racetrack coils wound with 100-m long IBS tapes produced by IEE. The quench current of the IBS coil at 10 T reached 81.25% of its quench current at self-field. In 2021, the IBS solenoid coil developed by IHEP reached 67 A at 30 T background field. The works verified the IBS conductor could be a promising candidate for the application in high field superconducting magnets.

R&D of high field model dipole is ongoing at IHEP, and in collaboration with related institutes working on fundamental sciences of superconductivity and the advanced HTS superconductors. A NbTi+Nb$_3$Sn twin-aperture magnet reached 12.47 T at 4.2 K in 2021. After that, Nb$_3$Sn+HTS (IBS or ReBCO) magnet with two $\Phi$ 45 mm apertures will be developed, aiming to reach 16+ T in 5 years, and 20~24 T in 10 years. The R&D will focus on the following key issues related with the high field superconducting magnet technology:

1) Explore new methods and related mechanism for HTS materials with superior comprehensive performance for applications. Reveal key factors in current-carrying capacities through studying microstructures and vortex dynamics. Develop advanced technologies of HTS wires for high field applications with high critical current density ($J_c$) and high mechanical strength.

2) Development of novel high-current-density HTS superconducting cables, and significant reduction of their costs. Exploration of novel structures and fabrication process of high field superconducting magnets, based on advanced superconducting materials and helium-free cooling method.

3) Exploration of novel stress management and quench protection methods for high field superconducting magnets, especially for high field insert coils with HTS conductors. Complete the prototype development with high field and $10^{-4}$ field quality, lay the foundation for the applications of advanced HTS technology in high-energy particle accelerators.

## 2.3 CEPC facilities, demonstration and industrialization

In CEPC TDR, EDR and construction phases, key technology development, measurement and test, etc. corresponding facilities are needed, such as SRF lab for cavity fabrication, surface treatment, vertical and horizontal test, cryomodule assembly, cooling down and operation with beam; high power klystron test; magnet measurement lab for long magnets (~5m long) measurement; vacuum chamber NEG coating, baking and test, etc. These facilities have been developed together with PAPS of IHEP's 4th generation 6GeV synchrotron light source, HEPS. As shown in Figs. 27 and 28.

HEPS as shown in Figs. 29 and 30 as a whole could be regarded as a system demonstration facility, which includes linac injector, booster and main ring, and the total cost of HEPS is more that 10% of CEPC CDR cost.

As a specific demonstration section, a CEPC mockup tunnel with key components have been designed including booster and collider ring magnets and SRF system as shown in Fig. 31 (left). To facilitate the industrialization, store, transportation, and installation logistics, a computer aided optimization software is under developing, as shown in Fig. 31 (right).

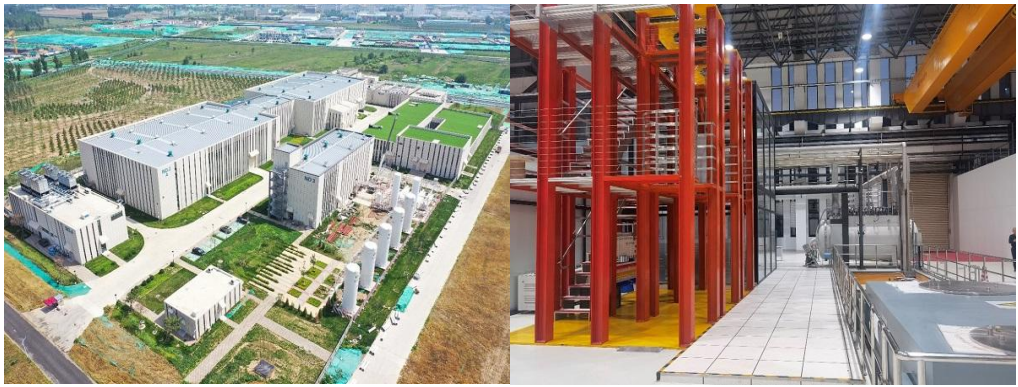

Fig. 27 CEPC SRF components, magnets and klystron measurement and development facility-PAPS of IHEP

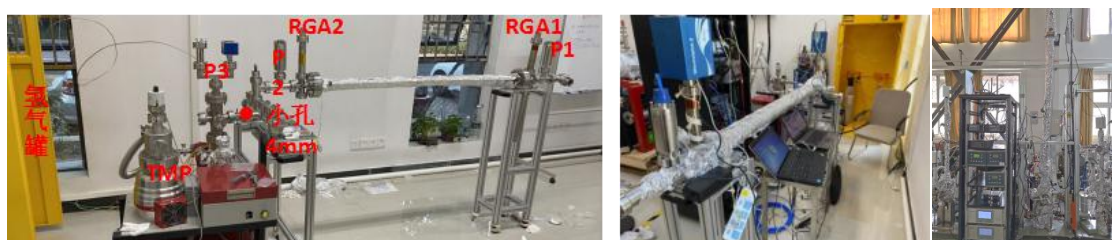

Fig. 28 CEPC vacuum system lab

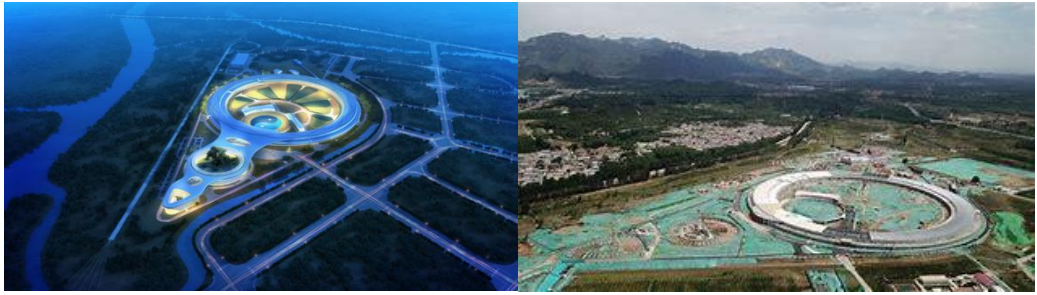

Fig. 29    HEPS, the 4th generation 6GeV light source of IHEP under construction to be completed in 2025, which has many common technologies with CEPC

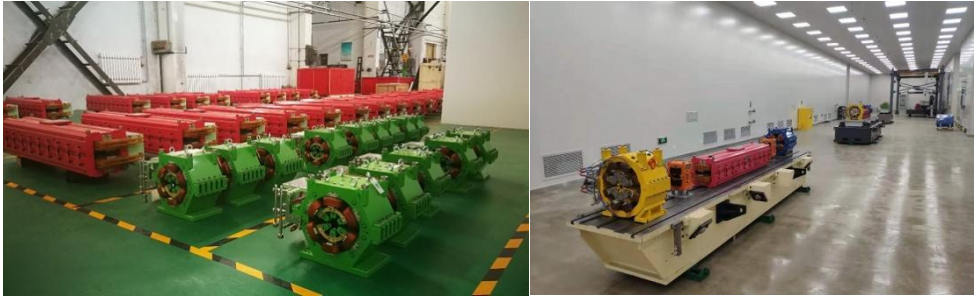

Fig. 30 HEPS magnets and sector alignment lab

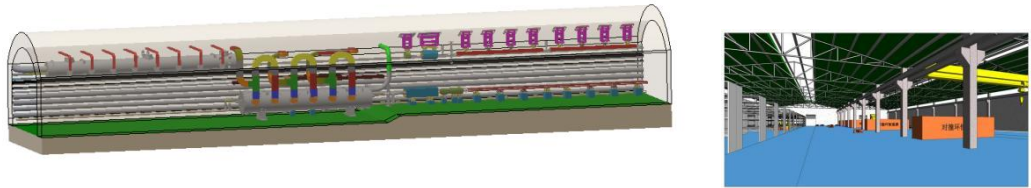

Fig. 31 CEPC mockup tunnel (left) and computer aided component stores and transportation logistic optimization design (right)

Since 2017, CEPC Industrial Promotion Consortium (CIPC) has been established and it has more than 70 members, as shown in Fig. 32. The working fields of the CIPC members have a wide spectrum, such as 650MHz high power high efficiency klystron, SRF cavity fabrication, normal and SC magnets, vacuum system, etc. International companies are also welcome to join CIPC.

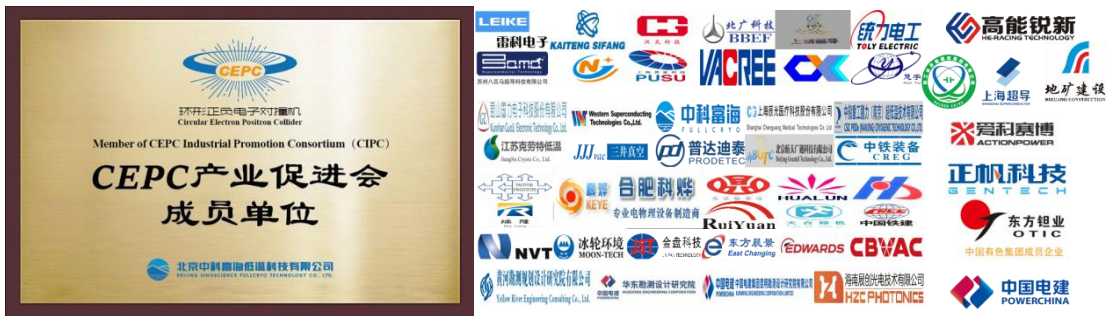

Fig. 32 CEPC Industrial Promotion Consortium (CIPC) and its more than 70 members

## 2.4 CEPC EDR and construction timeline (example)

After the CEPC TDR at the end of 2022, CEPC will enter into the phase of EDR (2023-2025):

-Engineering design of CEPC accelerator systems and components towards fabrication in an industrial way.

-Site selection converging to one or two sites with detailed feasibility studies (tunnel and infrastructures, environment).

-Site dependent civil engineering design implementation preparation.

-EDR document completed for government's approval of starting construction around 2026 (the starting of the "15th five year plan" of China).

In close collaboration with civil engineering companies, CEPC construction time line with civil engineering construction and installation is given in Fig. 33, and the components' mass production, storage and transportation for installation will be in a coordinated way to ensure the total construction schedule.

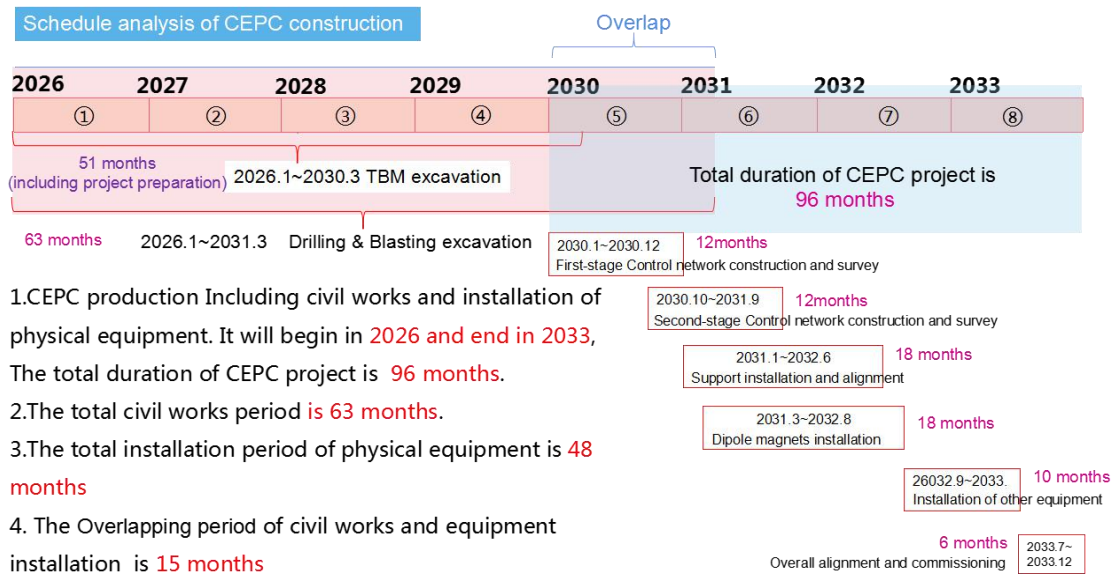

Fig. 33 CEPC civil construction example schedule

# 3. Staging Options and Upgrades

## 3.1 Energy upgrades

CEPC is a Higgs factory, and the first priority is to operate the machine at Higgs energy, followed by Z-pole and W energy runs. The norminal SR power per beam is 30MW for all energies.

CEPC could be upgraded to ttbar energy of 360GeV (center of mass), by increasing the SRF cavities and cryogenic system to increase VR voltage from 2.2GeV to 10GeV, as shown in Fig. 7 and the magnets in booster and collider rings have reserved margins to operate at 180GeV.

By constructing a Super proton proton Collider (SppC) in the CEPC tunnel,

collision energy in center of mass could read as high as 125TeV rang (see section 3.3).

## 3.2 Luminosity upgrades

CEPC luminosites at all energies could be upgraded by increasing the SR power per beam from 30MW at the Higgs, W, Z-pole and ttbar energies with luminosities $5\times10^{34}$ cm$^{-2}$s$^{-1}$ ,$16\times10^{34}$ cm$^{-2}$s$^{-1}$, $115\times10^{34}$ cm$^{-2}$s$^{-1}$ and $0.5\times10^{34}$ cm$^{-2}$s$^{-1}$ per interaction point, respectively, to 50MW SR power/beam at the Higgs, W, Z-pole and ttbar energies with luminosities $8.3\times10^{34}$ cm$^{-2}$s$^{-1}$, $27\times10^{34}$ cm$^{-2}$s$^{-1}$, $192\times10^{34}$ cm$^{-2}$s$^{-1}$ and $0.8\times10^{34}$ cm$^{-2}$s$^{-1}$ per interaction point, respectively, where the energy upgrade potential to tt-bar energy of 180GeV has been considered with luminosities of $0.5\times10^{34}$ cm$^{-2}$s$^{-1}$ and $0.8\times10^{34}$ cm$^{-2}$s$^{-1}$ correspongding to 30MW and 50MW beam SR power, respectively.

## 3.3 Experimental system upgrades and staging

As experimental system staging from CEPC, a Super proton proton Collider (SppC) could be installed in the same tunnel of CEPC without removing CEPC as shown in Fig. 34, and iron-based high-field superconducting magnets of at least 20T will be used to allow proton–proton collisions at a center-of-mass energy of 125TeV at a luminosity level of $4.3\times10^{34}$ cm$^{-2}$s$^{-1}$.

Electron proton collisions can also be realized in the CEPC-SPPC complex by bringing one beam from each of two colliders together and converting two *pp* collision IR for *e-p* collisions. The CM energy of *e-p* collision could reach 6.7TeV (by 62.5 TeV*p* × 180 GeV *e*). For 62.5TeV *p* × 120GeV *e* mode, the luminosity is $3.7\times10^{33}$ cm$^{-2}$s$^{-1}$ at one collision point [17].

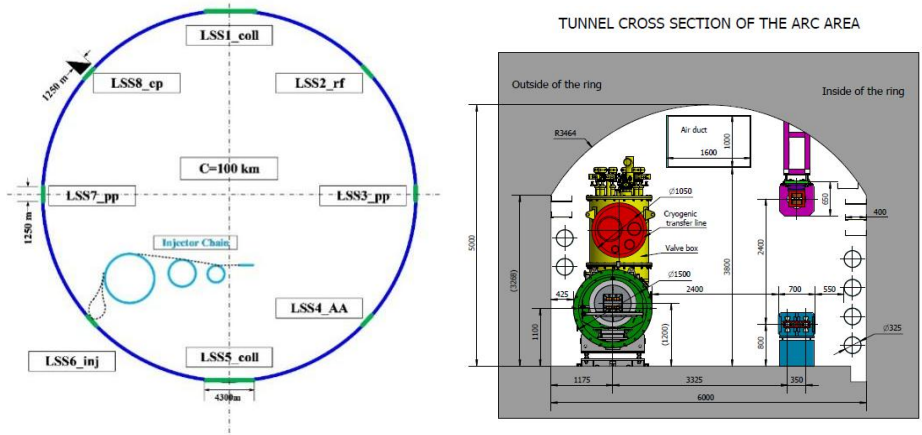

Fig. 34 SppC layout and CEPC-SppC in the same tunnel

# 4.Synergies with other concepts and/or existing facilities

## 4.1 Synergies on machine technologies

BEPCII (Beijing Electron Positron Collider) is a running collider at center of energy of 2-4.6GeV with luminosity around $0.7 \times 10^{33}$ cm$^{-2}$s$^{-1}$ at IHEP, and there is an upgrade plan of BEPC II in luminosity and center of energy up to 5.6GeV by increasing SRF voltage and a new SC QD0. There are many experiences can be drawn from BEPCII, which is a successor of BEPC put to operation in 1988.

Super KEK B is a modern electron positron collider in operation, where many technologies and experiences can be useful to CEPC, such as linac injector, damping ring, MDI, mini $\beta^*_y$ at IP, etc.

The 4th generation SR light source, HEPS, at IHEP under construction, which will be completed at the end of 2025, has many common technologies as CEPC, such as high precision magnets and vacuum system technology, etc. CEPC could also share the common facilities with HEPS, such as PAPS facility.

SHINE is a 8GeV CW X-FEL based on SC 1.3GHz linac technology, where SRF technologies, such as 12m long, 8 TESLA 9cell cavity cryomodules, large scale SRF accelerator system industrial mass production and operation, etc. SHINE is under construction in Shanghai, China, to be put into operation around 2026. SHINE will be very useful for CEPC SRF system development.

## 4.2 Synergies on detector technologies

CEPC has two detectors, and many advanced technologies will used. The synergies with ILC/CLIC detectors, such as ILD, is very useful and important. Collaborations with LHC and HL-LHC on detectors and corresponding contributions are very fruitful and important in CEPC detectors development and collaboration.

## 4.3 Synergies on conventional facilities and green power

Efficient utilization of energy in large scientific facilities is becoming one of the key issues for the sustainable development. Green energy and waste energy reuse technologies are under studies for future and existing projects, such as CEPC, ILC and LHC. Taking CEPC and ILC for examples, great efforts have been put to the development of 650MHz high power and high efficiency klystrons (around 80% of efficiency), and also on the high quality factor and high field SRF cavities.

## 4.4 Synergies for physics research

CEPC as an electron positron Higgs factory colliders have many common physics goals with ILC, CLIC, FCCee, and C3 linear collider proposed recently in USA [18]. LHC and HL-LHC physics research are very much complementary to those in lepton colliders.

# 5. Conclusions

CEPC as a Higss factory provides one of the future colliders for the high energy particle physics community and Sicence in general worldwide. CEPC has been firstly proposed by Chinese sicentists in Sept. 2012 just after the Higgs Boson discovery at CERN.With strong international and industrial participations, CEPC CDR has been completed in Nov. 2018, and accelerator TDR will be completed at the end of 2022. The key technologies of CEPC in collider/booster rings and linac injector have been intensively studied, and mastered in hands. CEPC will enter EDR phase in 2023 and will be completed at the end of 2025. CEPC team will work closely with Chinese central government, international/industrial collaborations, and the local host government in EDR phase towards the aim of putting CEPC into construction around 2026 (within the "15th " five year plan of China), and into operation around 2035.

# 6. Acknowledgements

Thanks go to CEPC-SppC accelerator team's hardworks, international and CIPC collaboration efforts.

Thanks go to the supports, encouragement, critical comments and suggestions from CEPC Institution Board(IB), Steering Committee(SC), International Advisory Committee (IAC)，and International Accelerator Review Committee( IARC).

Thanks go to the funds from Chinese MOST, NFSC, CAS and Scientist's Studio.